\documentclass[a4paper,fleqn,usenatbib]{mnras}

\usepackage{newtxtext,newtxmath}

\usepackage[T1]{fontenc}
\usepackage{ae,aecompl}

\usepackage{graphicx}	
\usepackage{amsmath}	
\usepackage{amssymb}	
\usepackage{pdfpages}
\usepackage{adjustbox}
\usepackage{hyperref}

\title[]{New constraints on the initial parameters of low-mass star formation from chemical modeling}

\author[T. Vidal et al.]{
Thomas H. G. Vidal$^{1}$\thanks{E-mail: thomas.vidal@u-bordeaux.fr},
Pierre Gratier$^{1}$,
Neil Vaytet$^{2}$,
Audrey Coutens$^{1}$,\newauthor
and Valentine Wakelam$^{1}$
\\
$^{1}$Laboratoire d'astrophysique de Bordeaux, Univ. Bordeaux, CNRS, B18N, allée Geoffroy Saint-Hilaire, 33615 Pessac, France\\
$^{2}$Data Management and Software Centre, the European Spallation Source ERIC, Ole Maaløes Vej 3, 2200, Copenhagen, Denmark\\
}

\date{Accepted XXX. Received YYY; in original form ZZZ}

\pubyear{2017}

\begin{document}
\label{firstpage}
\pagerange{\pageref{firstpage}--\pageref{lastpage}}
\maketitle

\begin{abstract}
The complexity of physico-chemical models of star formation is increasing, with models that take into account new processes and more realistic setups. These models allow astrochemists to compute the evolution of chemical species throughout star formation. Hence, comparing the outputs of such models to observations allows  to bring new constraints on star formation.\\
The work presented in this paper is based on the recent public release of a database of radiation hydrodynamical low-mass star formation models. We used this database as physical parameters to compute the time dependent chemical composition of collapsing cores with a 3-phase gas-grain model. The results are analyzed to find chemical tracers of the initial physical parameters of collapse such as the mass, radius, temperature, density, and free-fall time. They are also compared to observed molecular abundances of Class 0 protostars.\\
We find numerous tracers of the initial parameters of collapse, except for the initial mass. More particularly, we find that gas phase CH$_3$CN, NS and OCS trace the initial temperature, while H$_2$CS trace the initial density and free-fall time of the parent cloud. The comparison of our results with a sample of 12 Class 0 low mass protostars allows us to constrain the initial parameters of collapse of low-mass prestellar cores. We find that low-mass protostars are preferentially formed within large cores with radii greater than 20000 au, masses between 2 and 4 M$_{\odot}$, temperatures lower or equal to 15 K, and densities between 6$\times10^4$ and 2.5$\times10^5$ part.cm$^{-3}$, corresponding to free-fall times between 100 and 200 kyrs.
\end{abstract}




\section{Introduction}

It is now commonly accepted that low-mass stars form from gravitational collapses within dense molecular clouds \citep[see][for a review]{McKee08}. Indeed, the physical processes that govern the dynamic of these dense clouds, such as self-gravity, radiation field, and magnetic field, cause density fluctuations that allow the formation of denser regions within the clouds, called prestellar cores. When the thermal pressure within a prestellar core can no longer compensate its internal gravitational force, it starts collapsing onto itself. The prestellar core is initially optically thin and collapses isothermally, with compressional heating evacuated through radiation. The efficiency of this radiative cooling drops as the density of the core increases at its center. When it can no longer equilibrate with the compressive heating, the evolution of the matter at the center of the core becomes adiabatic, and a hydrostatic core, known as the first hydrostatic core (FHSC), or first Larson's core \citep{Larson69}, is formed at the center. The protostar is born. During a short period of time \citep[< 1000 years, see for example][]{Vaytet17}, the FHSC will continue to accrete surrounding material from its envelope, consequently increasing its mass, density and temperature. When its temperature exceeds $\approx$2000 K, the dissociation of H$_2$ molecules inside the core is triggered. Because of the endothermic nature of this process, a second phase of collapse begins. When most of the H$_2$ molecules have been split, the second collapse ceases and a much more dense and compact core forms, called the second Larson's core \citep{Larson69,Masunaga00,Vaytet13}. The core continues to slowly accrete material for a few hundred thousand years, until its temperature is high enough to trigger nuclear fusion reactions and the protostar becomes a young star.\\

In order to put constraints on the chemical evolution of the interstellar matter, from the dense molecular clouds to the multiple components of stellar systems, chemical and physical odels of prestellar collapse need to be associated. Such studies have already been conducted, from simple one-dimensional models \citep{Ceccarelli96,Doty04,Aikawa08}, to multi dimensional ones \citep{Visser11,Furuya12,Hincelin16}. Moreover, the complexity of these models keep on increasing. Indeed, physical models of collapse can now include several physical processes in increasingly realistic setups \citep[see for instance][]{Commercon10,Li14,Tomida15,Matsumoto15}. Chemical models are also constantly improved, notably via the implementation of multi-phases modeling \citep{Taquet12,Ruaud16} (introducing a difference between the most external layers of the ices on the grains and the rest of the bulk ice) and new chemical processes \citep{Garrod07,Ruaud15}, but also thanks to public comprehensive chemical database such as \textsc{UMIST} \citep{McElroy13} and \textsc{KIDA} \citep{Wakelam15}. Such refinement of physico-chemical model allows to converge towards a more realistic representation of star formation as well as a better interpretation of the observations of star forming regions.\\

In this context, the work presented in this paper was initially motivated by the recent public release of a database of radiation hydrodynamical (RHD) low-mass star formation models by \citet{Vaytet17} (\url{http://starformation.hpc.ku.dk/grid-of-protostars}). Each model simulates the 1D collapse of a spherical dark cloud described by a given set of initial physical parameters, and stops shortly after the formation of the second Larson's core, when the collapse is believed to cease. The idea is to combine the physical models of this database with the \textsc{Nautilus} chemical model in order to derive useful information on the physical parameters of low-mass star formation. More particularly, it aims to challenge the possibility of putting constraints on the initial parameters of collapse of dark clouds using chemical modeling. In the first section, we detail the dataset of \citet{Vaytet17} and the chemical model. In the second section, we detail how both were combined to obtain a set of physico-chemical models. From the results of these models, we highlight species that could be possible tracers of initial physical parameters of collapse (hereafter IPPC). Additionally, we developed a simple method to find constraints on the IPPC of observed low-mass Class 0 protostars. These results are presented respectively in the third and fourth sections. We will then discuss these results to finally conclude.\\

\section{The physical and chemical models}

In order to compute the time dependent chemical composition of collapsing cores, we have used the physical structures computed by a 1D RHD model by \citet{Vaytet17}. This model computes the collapse of a spherical cloud in a Lagrangian approach. The radius of the sphere is split in independent cells of materials that will move towards the center during the collapse, while their respective temperatures and densities will increase. Hence, for each cell and at each time step of the model, its radius, temperature, and density are known. These parameters are then used as inputs for the chemical model. In the following are described the RHD models database as well as the chemical model we used in this paper.  

\subsection{The RHD models database}

\begin{figure*}
        \begin{center}
                \includegraphics[scale=0.4]{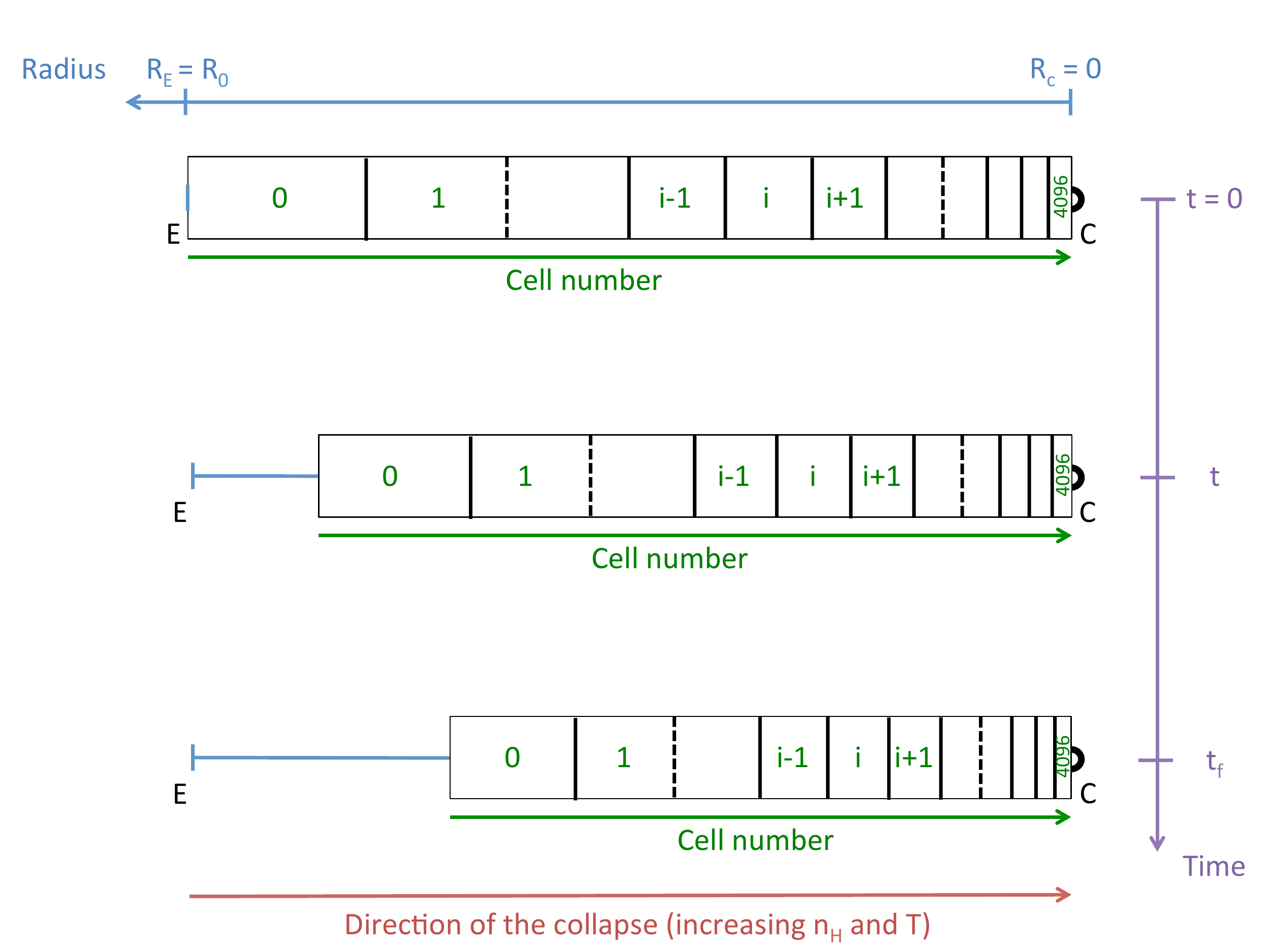}
                \caption{Representation of the one-dimensional Lagrangian grid used in the RHD models.}
                \label{fig_31a}
        \end{center}
\end{figure*}

The dataset of 1D low-mass star formation models from \citet{Vaytet17} is composed of 143 models that follow the formation of a protostar, from the initial isothermal collapse to the FHSC, and then to the second collapse and the formation of the second Larson's core. To solve the equation of RHD for each model, \citet{Vaytet17} used a 1D fully implicit Godunov Lagrangian code with a grid comprising 4068 cells. Figure \ref{fig_31a} displays for a given core a representation of the mesh used in the model and its evolution throughout the collapse. In the initial state ($t=0$, top panel), the mesh is superimposed on the radius of the core. Hence the outer radius of the outermost cell is equal to the radius R$_0$ of the initial BE sphere, and the inner radius of the innermost cell is $R_c = 0$. Moreover, in order to obtain a higher resolution towards the center of collapse C (black cercle), the cell sizes are decreased progressively with decreasing radius such as:

\begin{eqnarray}
	\Delta R_{i} = (1+\alpha)\Delta R_{i+1}
\end{eqnarray}

where $\Delta R_{i}$ is the width of cell $i$ and $\alpha = 8\times10^{-4}$. During the collapse (t, middle panel), the mesh moves toward the center of the core, each of the cells enclosing the mass of the parcel of material it was initially matched with, notably by decreasing in size. Indeed, since the mass of each cell stays constant during the collapse, their respective densities can increase only if their sizes decrease. In the following, I will therefore use the term "cell" to refer to its corresponding parcel of material.\\

Each of the models is defined by a set of initial parameters and considers the initial parent cloud as a Bonnor-Ebert (BE) sphere \citep{Bonnor56,Ebert55}. For all models the value of the density contrast of the initial BE density profile is $\gamma = \rho_c/\rho_0 = 14.1$, which is the maximal value to obtain a marginally stable sphere, and which corresponds to a dimensionless radius $\xi = 6.45$. In order to fully define the gas density of each initial BE sphere, fixed values for its temperature $T_0$ and radius $R_0$ were chosen. For these parameters, the central density and mass of the corresponding critical BE sphere, \textit{i.e.} which defines a stable cloud, can be found via:

\begin{eqnarray}
	\rho_c &=& \Big(\frac{\xi}{R_0}\Big)^2\frac{k_BT_0}{4\pi\mu m_HG} \label{radiuscorr}\\
	M_{\text{BE}} &= &2.4\frac{c_s^2}{G}R_0
\end{eqnarray}

where $k_B$ is the Boltzmann's constant, $\mu$ the mean atomic weight (=2.31) and $m_H$ the hydrogen atom mass. The initial cloud mass $M_0$ is then chosen for the ratio $\epsilon=\frac{M_{\text{BE}}}{M_0}$ to have values below 1, corresponding to a cloud that would undergo gravitational collapse. The central density of the cloud can now be expressed as:

\begin{eqnarray}
	\rho_c = 5.78M_0\Big(\frac{4\pi}{3}R_0^3\Big)^{-1}
	\label{density}
\end{eqnarray}

And its corresponding free-fall time as:

\begin{eqnarray}
	t_{\text{ff}} = \sqrt{\frac{3\pi}{32G\rho_c}}
	\label{tff}
\end{eqnarray}

The initial parameter space (cf Fig. 1 of \citet{Vaytet17}) was chosen for $\epsilon$ to cover a wide range of values between 0 and 1 while using a set of different initial cloud masses, temperatures and radii, respectively ranging from 0.2 to 8 M$_{\odot}$, 5 to 30 K and 3000 to 30000 au. Such a wide parameter space allows to work with models covering a large diversity of molecular clouds in which low-mass protostars can form. The second column of table \ref{tab_31} (labelled "Original dataset") summarizes the range of values for each IPPC of interest in the present study. For the sake of clarity, it should be noted that we include the free-fall time in the term "IPPC", even if it is not a physical parameter.\\

\begin{table}
\caption{Initial physical parameters ranges for the original dataset and the reference dataset for chemical modeling }
	\begin{center}
		\begin{tabular}{l c c}
		\hline
		\hline
   		IPPC & Original dataset & Reference dataset \\
   		\hline
		\hline
		$M_0$ (M$_{\odot}$)            &  [0.2 ; 8]       & [0.5 ; 8] \\
   		$T_0$ (K)                            & [5 ; 30]     & [10 ; 30] \\
		$R_0$ (au)                         & [3000 ; 30000]       &  [3000 ; 30000] \\
		$\rho_0$ (part.cm$^{-3}$)   & [2.98$\times10^4$ ; 1.20$\times10^8$]   &  [5.98$\times10^4$ ; 1.20$\times10^8$]  \\
		$\rho_c$ (part.cm$^{-3}$)   & [4.20$\times10^5$ ; 1.69$\times10^9$]   &  [8.43$\times10^5$ ; 1.69$\times10^9$]  \\
		$t_{\text{ff}}$ (kyrs)    & [4.31 ; 272.83]  &  [4.31 ; 192.92] \\
		\hline
		\hline
 		\end{tabular}
	\end{center}
  	\label{tab_31}
\end{table}

\subsection{The chemical model}

The chemical model we use is the \textsc{Nautilus 1.1} chemical model, which computes the evolution of chemical abundances for a given set of physical and chemical parameters. It can simulate a three-phase chemistry including gas-phase, grain-surface and grain bulk chemistry, along with the possible exchanges between the different phases \citep{Ruaud16}. These exchanges are:  the adsorption of gas-phase species onto grain surfaces, the thermal and non-thermal desorption of species from the grain surface into the gas-phase, and the surface-bulk and bulk-surface exchange of species. The chemical desorption process used in the model is the one depicted in \citet{Garrod07} where we consider a 1\% efficiency evaporation for all species. Moreover, the grain chemistry takes into account the standard direct photodissociation by photons along with the photodissociation induced by secondary UV photons \citep{Prasad83}. These processes are effective on the surface as well as in the bulk of the grains. The model also takes into account the newly implemented competition between reaction, diffusion, and evaporation as suggested by \citet{Chang07} and \citet{Garrod11}. The diffusion energies of each species are computed as a fraction of their binding energies. We take for the surface a value of this ratio of 0.4 as suggested by experiments and theoretical work made on H \citep[see][and reference therein]{Ruaud16}, CO and CO$_2$ \citep[see][]{Karssemeijer14}. This value is then extrapolated to every species on the surface. For the bulk, we take a value of 0.8 \citep[see also][]{Ruaud16}. Given the high temperature regimes encountered in this study, we use the \textit{ad hoc} formation mechanism for H$_2$ described in \citet{Harada10}, as depicted in \citet{Vidal18}.\\

The reference chemical network is \textit{kida.uva.2014} \citep[see][]{Wakelam15} for the gas-phase and the one described in \citet{Ruaud16} for the grains. To this was added the sulphur network detailed in \citet{Vidal17} \citep[including the reactions given in][]{Druard12}, as well as the chemical schemes for carbon chains proposed in \citet{Wakelam15b}, \citet{Loison16}, \citet{Hickson16} and \citet{Loison17}. Note that all abundances in this paper are expressed with respect to the total H density.

\section{Pre-treatment of the data}

\subsection{Selection of the reference dataset of physical models} \label{selecmodel}

For one set of initial parameters, \textit{i.e.} one model, the RHD code gives the physical conditions of the 4096 cells of the Lagrangian grid at every time step. Detailed examples of the cell evolution in the density-temperature plane are presented in Figure 3 of \citet{Vaytet17}. We first select the models and their respective cells in order to satisfy the temperature validity domain of [10 K - 800 K] of the chemical network. With this first selection, 110 models out of 143 are useable for the chemical model application. Plus, for the computing time of \textsc{Nautilus} to be acceptable, we selected a number of 16 cells per RHD model between the outermost cell of the grid and the last inner cell which temperatures stay below 800 K, selecting one cell every $4096/16 = 256$ cells, including the outermost one. As such the reference dataset for chemical modeling includes 110 RHD models defined by a set of initial parameters, each of them comprising 16 cells. \\

In the third column of table \ref{tab_31} (labeled "Reference dataset") are displayed the resulting ranges of the selected IPPC. By comparing them with those of the original dataset, we can see that the selection process has only a small impact on the ranges of the IPPC. Indeed, except for $T_0$, the differences of range for $\rho_0$ and $t_{\text{ff}}$ are due to only two models that are ruled out because of their initial temperatures. Hence, the reference dataset still accounts for a significant range of possible IPPC.

\subsection{Initial parameters of the chemical modeling} 

After selecting the reference dataset of physical models, the idea is now to run for each cell of the selected clouds a 0D dynamic chemical model while separately saving its radius at each time step.\\

As dark clouds are believed to evolve during a period of time which is not negligible for the chemistry before they can become gravitationally unstable, we first had to consider the chemical evolution of each cloud before its collapse to obtain chemical input data for the corresponding chemical model. To that effect, a standard 0D model of a dark cloud was run for each of the 110 selected clouds, using as temperature their respective initial temperatures $T_0$ and as proton total density the initial external density of the collapsing cloud ($n_H = \rho_0$). The choice to use the initial external density and not the central density can be justified by the fact that, given the selection of cells explained in the previous section, most of the selected cells have a initial density closer to the external density of their respective clouds. Other parameters of interest for the chemical modeling such as visual extinction and cosmic ionization rate were set to their commonly used values for dark cloud models, respectively 15 mag and $1.3\times10^{-17}$ s$^{-1}$. The set of initial abundances used for all pre-collapse dark cloud runs is summarized in table \ref{tab_31b}. The resulting models were run for 10$^{6}$ years, and their computed chemical compositions used as inputs for the chemical modeling of the corresponding cells of collapsing clouds.\\

\begin{table}
\caption{Initial abundances and cosmic-ray ionization rate $\zeta$ (in s$^{-1}$). *$a(b)$ stands for $a\times10^b$.}
	\begin{center}
		\begin{tabular}{l r r}
		\hline
		\hline
   		Element & $n_i/n_H$* & References \\
   		\hline
		H$_2$    & 0.5             &    \\
   		He          & 0.09           & 1 \\
		N            & 6.2(-5)       &  2 \\
		O            & 2.4(-4)       &  3 \\
		C$^+$    & 1.7(-4)       &  2 \\
		S$^+$    & 1.5(-5)       &  2 \\
		Si$^+$   & 8.0(-9)       &  4 \\
		Fe$^+$  & 3.0(-9)       & 4\\
		Na$^+$  & 2.0(-9)       & 4 \\
		Mg$^+$ & 7.0(-9)       & 4\\
		P$^+$   & 2.0(-10)     & 4\\
		Cl$^+$  & 1.0(-9)       & 4 \\
		F           & 6.7(-9)     & 5 \\
		\hline
		$\zeta$ & $1.3\times 10^{-17}$ & \\
		\hline
 		\end{tabular}
	\end{center}
	\medskip{(1) \citet{Wakelam08}, (2) \citet{Jenkins09}, (3) \citet{Hincelin11}, (4) Low-metal abundances from \citet{Graedel82}, (5) Depleted value from \citet{Neufeld05}}
  	\label{tab_31b}
\end{table}

In the following, we treat all the selected cells independently from each other. Hence, the outputs of the chemical modeling of the collapsing clouds describe for each of the $110\times16 = 1760$ cells their respective chemical evolution during the collapse. Because most of the youngest low-mass stellar objects studied in the literature are Class 0 protostars, we focus the study of these outputs on the final time step of the chemical models, when both Larson's cores are formed, and the object is therefore a Class 0 protostar.  

\subsection{Identification of two chemical regimes} \label{regime}

In order to find possible tracers of the IPPC, we first studied the correlation plots between the abundance distribution of each species at the final time (when the protostar is formed) and the IPPC ($M_0$, $T_0$, $R_0$, $\rho_0$ and $t_{\text{ff}}$). We did not find any obvious correlations with this method. However, two chemical regimes seemed to appear in the abundance distributions of most of the species: a low temperature regime (T $\leq$ 100 K) where most species are depleted onto grains, and a high temperature regime (T $\geq$ 145 K), dominated by the evaporation of species from the grain mantles. In order to visualize these temperature intervals, we plot in figure \ref{fig_33} the abundance of water for all cells as a function of their final temperatures and radii. Cells within the first temperature interval are located in the outer part of the clouds (further out than a few tens of au), \textit{i.e.} their Envelope Region, while cells within the two remaining intervals are located in the inner part of the clouds, \textit{i.e.} their Hot Corino Region. we can thus define regions corresponding to each temperature range:

\begin{figure}
        \begin{center}
	        \includegraphics[scale=0.45]{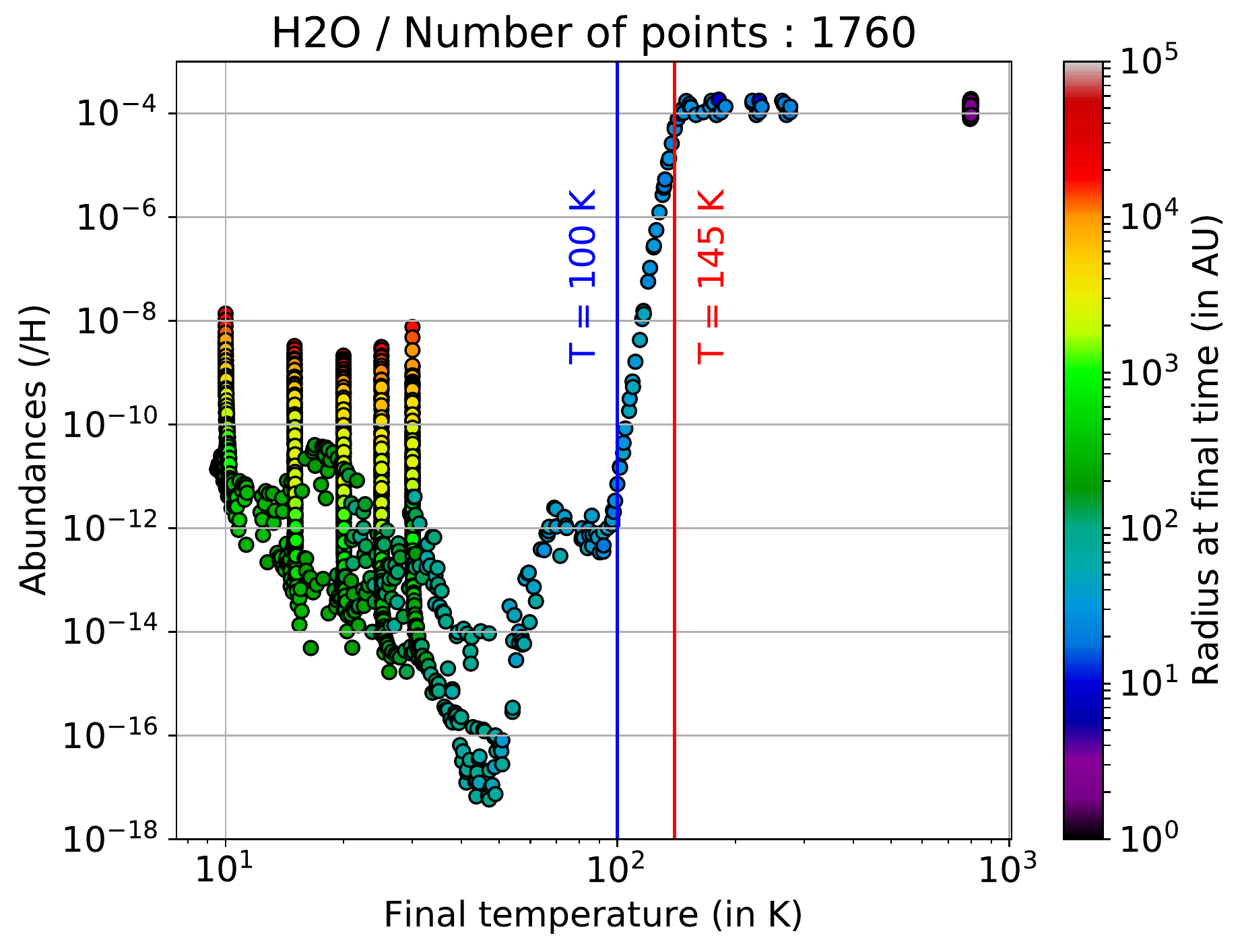}
                \caption{Distribution of the abundances of H$_2$O relative to H at the final time of the collapse as a function of the final temperature. The color coding represents the final radii of the cells. The blue vertical bar represents the upper temperature limit on the Envelopes Region, and the red one the lower temperature limit on the Hot Corinos Region.}
                \label{fig_33}
        \end{center}
\end{figure}

\begin{enumerate}
\item An Envelopes Region (ER), which corresponds to the temperature interval of 10 to 100 K and contains 1590 cells,\\
\item A Transitional Region (TR), which corresponds to the temperature interval of 100 to 145 K and contains 32 cells,\\
\item A Hot Corinos Region (HCR), which corresponds to the temperature interval above 145 K and contains 138 cells.
\end{enumerate}

It should be noted that the separation of the cells regarding final temperatures studied here for water is common to most of the species of the chemical network. However, for the remaining species, the defined temperature intervals will not necessarily fit their abundance distributions. Indeed, TPD experiments on more volatile species ($E_D < E_D^{H_2O}$) show that they begin to desorb in the gas phase before water, but have their main desorption peak at the water desorption temperature as they co-desorb with water which is the main ices constituant \citep[see][]{Fayolle11,Ruaud16}. Plus, more refractory species ($E_D > E_D^{H_2O}$) partly co-desorb with water, but are expected to desorb mainly at temperature higher than water \citep[see for example][]{Chaabouni18}. However, as the defined intervals would be consistent in first approximation with more volatile species, and because more refractory species only account for less than 15\% of the total number of species in the chemical network, we choose to apply the region differentiation as defined above to all species regardless of their desorption energies. For the following correlation study, we decide to ignore the 33 points from the TR because they represent less than 2\% of the total number of cells. Hence, the resulting loss of information is negligible.\\

\section{Search for tracers of the IPPC}\label{tracers}

We study the correlations between species abundances and IPPC using the Spearman's correlation coefficient ($\rho_S$). It assesses how well the relationship between two variables can be described using a monotonic function. Hence, it is more suitable to this study than the commonly used Pearson's coefficient, which evaluates the linear relationship between two variables. Indeed, we first used the latter and found that linear relationships between the abundance of a given species and the IPPC were scarce. Finally we restrain our study to observed or observable species. \\

Note that in the ER, the results of correlations are inconclusive since the majority of the resulting correlation coefficients are too low (< 0.2 in absolute value) or concern unobservable species. This result is not conflicting with the use of the ER dataset to constrain the IPPC as detailed in section \ref{compobs}. Therefore, we only present here the correlation results for the HCR.\\

\begin{table}
\caption{Summary of the calculated Spearman's coefficients between the abundances at final time of the 482 gaseous species of the chemical network and the IPPC, using the 138 cells of the Hot Corino Region.}
	\begin{center}
		\begin{tabular}{l c c }
		\hline
		\hline
   		Initial physical parameters & $|\rho_S|_{max}$   & Number of species with $|\rho_S|>0.6$\\
   		\hline
		\hline
		$M_0$      &   0.389   &  0\\
   		$T_0$        &   0.978  & 137 \\
		$R_0$        &  0.909   & 88 \\
		$\rho_0$   & 0.983    &  117  \\
		$t_{\text{ff}}$  & 0.983  & 117  \\
		\hline
		\hline
 		\end{tabular}
	\end{center}
  	\label{tab_32}
\end{table}

Table \ref{tab_32} displays maximum values of the Spearman's coefficient obtained for each IPPC, $|\rho_S|_{max}$, as well as the number of species linked with them, \textit{i.e.} with $|\rho_S|>0.6$. In the HCR, it appears that numerous correlations can be found between species abundances and IPPC, except for M$_0$ for which we find only low Spearman coefficients ( < 0.4 in absolute value).\\

Among the species linked with one of the IPPC, 11 are detected toward the well studied Class 0 protostar IRAS 16293-2422. Here we present the results of the correlation for 4 of these species, for which the abundances of the HCR cells span over more than 3 orders of magnitude: CH$_3$CN, H$_2$CS, NS, and OCS. For the other species for which the abundances span over a smaller range of values, the results would be difficult to compare with observations. In that case, the variation of abundances would indeed not be significant enough for the corresponding species to be clear tracers, especially given the common uncertainties on observed abundances.\\

Figures \ref{fig_33a}, \ref{fig_34}, \ref{fig_35}, and \ref{fig_36} display the respective abundances of the 4 selected species as a function of their respective best correlated IPPC. H$_2$CS is best correlated with the free-fall time and the density of the collapsing clouds, when CH$_3$CN, NS and OCS are best correlated with the initial temperature. Here we try to understand how these correlations can be explained with chemistry.\\

\subsection{CH$_3$CN}

\begin{figure}
        \begin{center}
	        \includegraphics[scale=0.45]{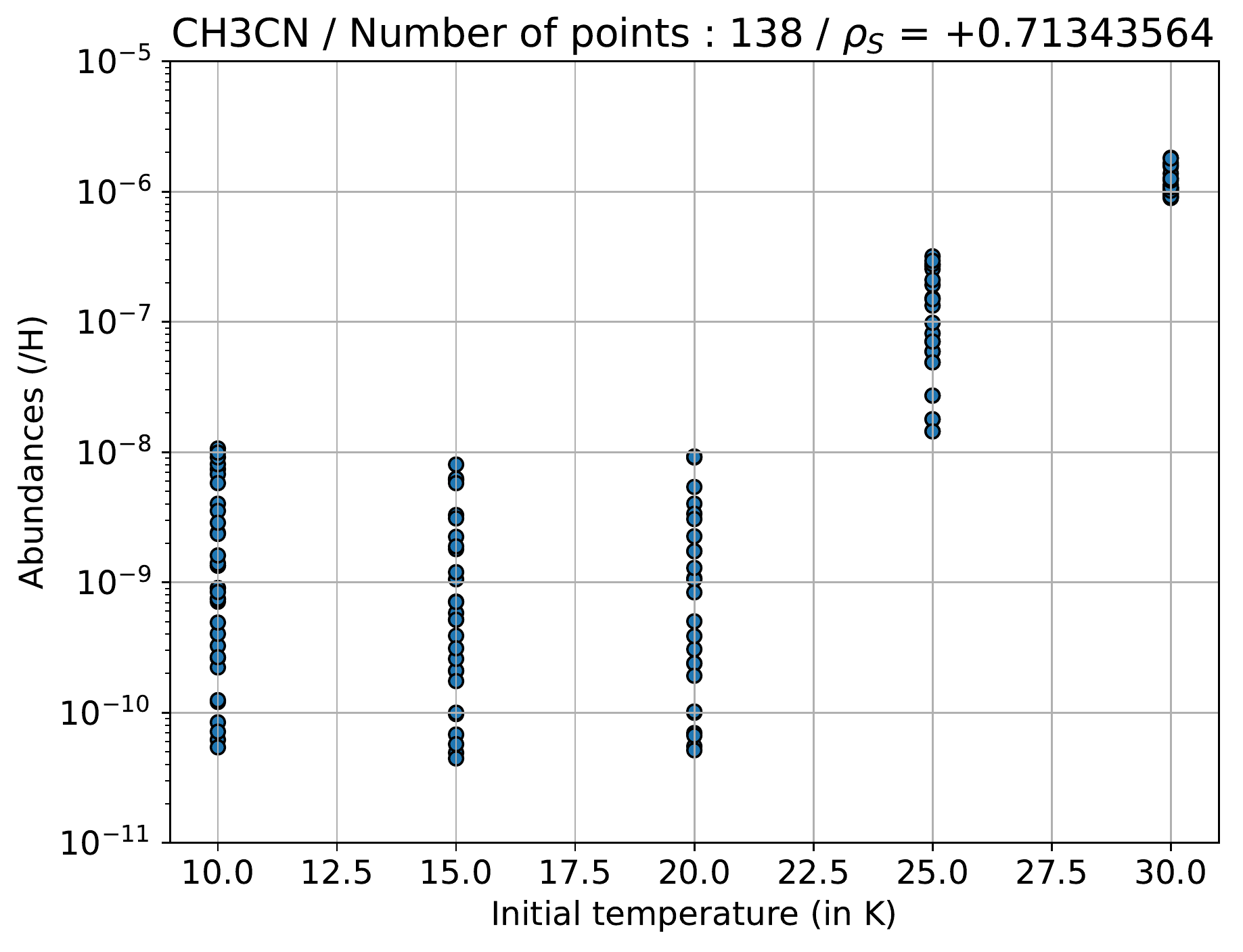}
                \caption{Distribution of the abundances of CH$_3$CN relative to H at the final time of the collapse as a function of the initial temperature.}
                \label{fig_33a}
        \end{center}
\end{figure}

Figure \ref{fig_33a} shows that the abundance of acetonitrile (CH$_3$CN) in the HCR is positively correlated with the initial temperature of the collapsing clouds with a Spearman's coefficient of 0.713. This means that the more abundant CH$_3$CN is in a hot corino, the higher its parent cloud's initial temperature. More precisely, an observer that would detect CH$_3$CN in a hot corino with an observed abundance $[CH_3CN]_{\text{obs}}$ could interpret this correlation such as:

\begin{itemize}
\item If $[CH_3CN]_{\text{obs}} \leq 10^{-8}$, $T_0 \leq 20$ K,\\
\item If $10^{-8} \leq [CH_3CN]_{\text{obs}} < 10^{-6}$, $T_0 \approx 25$ K,\\
\item If $[CH_3CN]_{\text{obs}} \geq 10^{-6}$, $T_0 \geq 30$ K.
\end{itemize}

Indeed, in the chemical model, CH$_3$CN is known to efficiently form on the grain surfaces during the dark cloud lifetime \citep[see][]{2018MNRAS.481.5651A} through the hydrogenation of adsorbed H$_2$CCN, the latter being preferentially formed in the gas phase following:

\begin{eqnarray}
	\text{CN} + \text{CH$_3$} \to \text{H} + \text{H$_2$CCN} \label{eq_3}
\end{eqnarray}

At the end of the parent dark cloud runs, CH$_3$CN abundance reaches higher values when the temperature of the cloud is higher. This increase can go up to 2 orders of magnitude in the interval of initial temperatures considered, which would explain the shape of the abundance distribution of CH$_3$CN seen on figure \ref{fig_33a}, and the resulting correlation. This increase in CH$_3$CN abundance with the temperature of the dark clouds is due to the fact that higher temperatures favor the diffusion of grains surface species, which allows reactions inefficient at lower temperatures to become efficient. In the case of CH$_3$CN, it is the reaction:

\begin{eqnarray}
	\text{s-CN} + \text{s-CH$_3$} \to \text{s-CH$_3$CN}
\end{eqnarray}

that becomes efficient because of the increased diffusivity of solid CN and CH$_3$, resulting in the increase in CH$_3$CN abundance.

Afterwards during the collapse, CH$_3$CN thermally desorbs in the gas phase where its abundance remains stable because its main destruction reaction is:

\begin{eqnarray}
	\text{CH$_3$CN} + \text{H$_3$O$^+$} \to \text{H$_2$O} + \text{CH$_3$CNH$^+$} 
\end{eqnarray}

 which is not very efficient because of the low abundance of H$_3$O$^+$. Moreover, its product CH$_3$CNH$^+$ forms CH$_3$CN again via electronic recombination. 

\subsection{H$_2$CS}

\begin{figure*}
        \begin{center}
	        \makebox[\textwidth]{\includegraphics[scale=0.45]{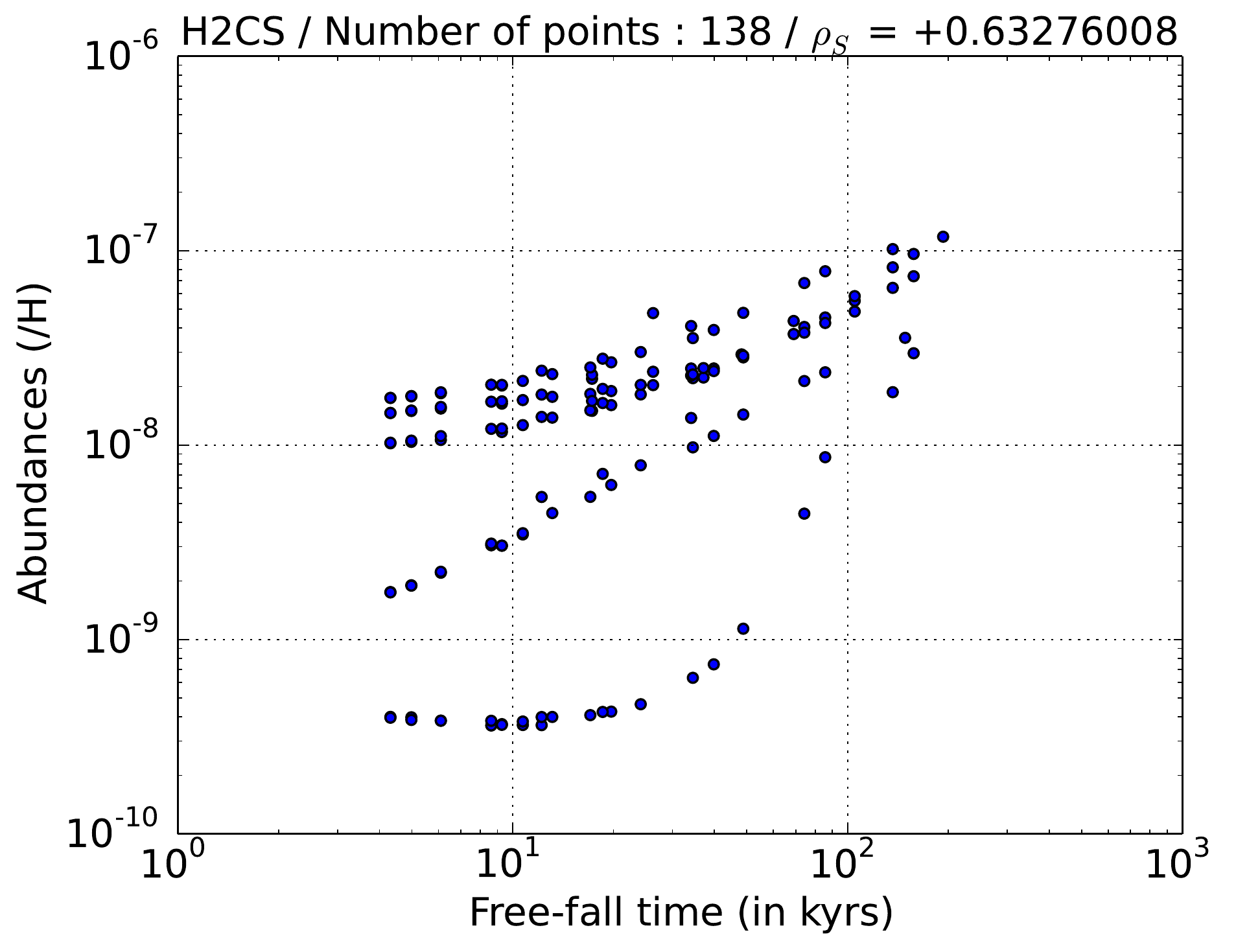}
	        \includegraphics[scale=0.45]{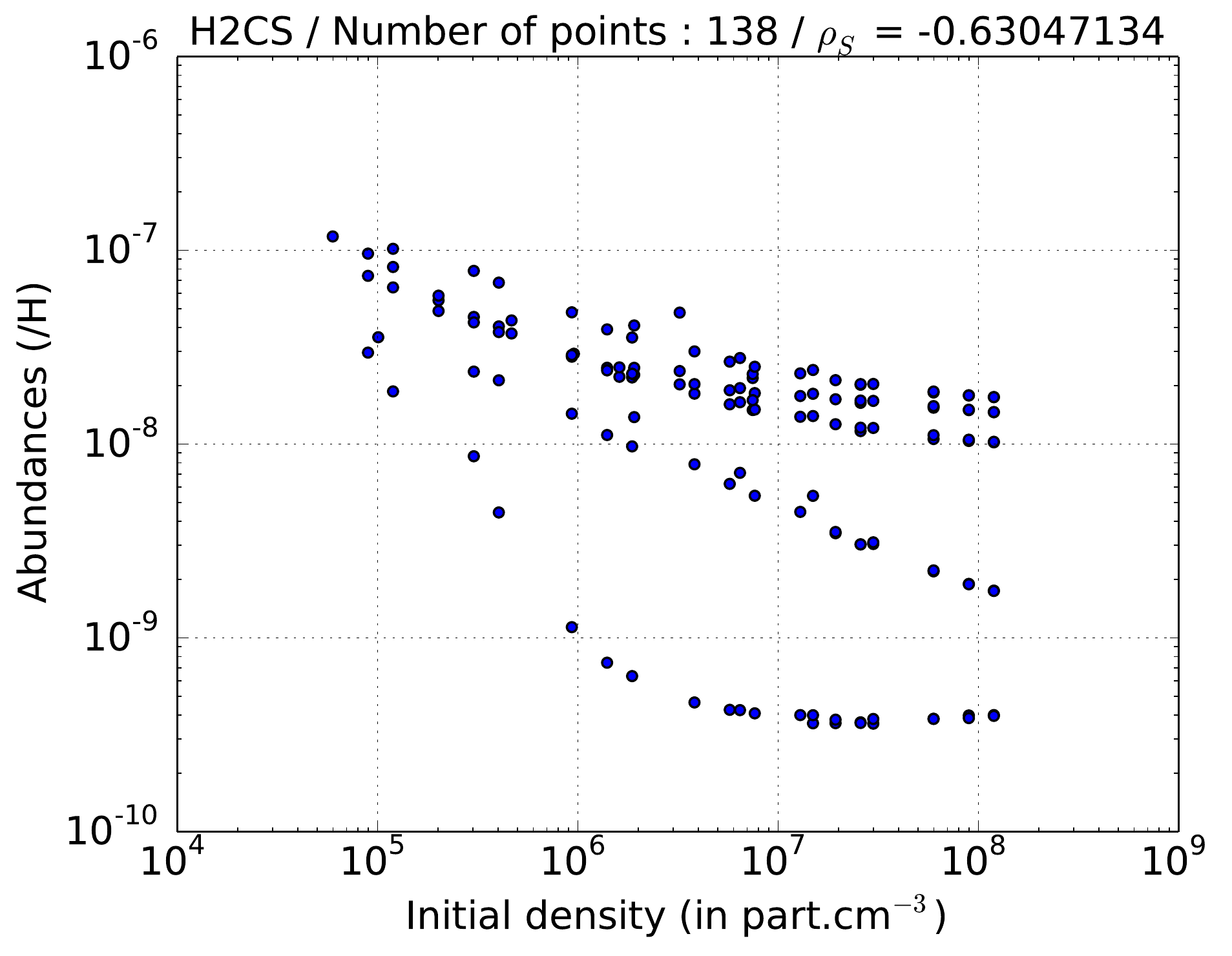}}
                \caption{Distribution of the abundances of H$_2$CS relative to H at the final time of the collapse as a function of the free-fall time (left panel) and the initial density (right panel).}
                \label{fig_34}
        \end{center}
\end{figure*}

Figure \ref{fig_34} shows that the abundance of thioformaldehyde (H$_2$CS) in the HCR is positively correlated with the free-fall time with a Spearman's coefficient of 0.633 (left panel), as well as anti-correlated with the initial density with a Spearman's coefficient of -0.630 (right panel). Indeed, because $t_{\text{ff}} \propto \frac{1}{\sqrt{\rho_c}}$ (see equation \ref{tff}), parent clouds with a low initial density have long free fall time and vice-versa. Given an observed abundance $[H_2CS]_{\text{obs}}$ in a hot corino, an observer could interpret this correlation such as:

 \begin{itemize}
\item If $[H_2CS]_{\text{obs}}$ is close to 10$^{-7}$, then the free-fall time of the parent cloud of the hot corino would be rather long, around 10$^2$ kyrs. Respectively, the initial density would be around 10$^5$ part.cm$^{-3}$,\\
\item If $[H_2CS]_{\text{obs}} < 10^{-9}$, then the free-fall time would be rather short, below 50 kyrs. Respectively, the initial density would be above 10$^6$ part.cm$^{-3}$
\end{itemize} 

We find that in parent clouds with high density, H$_2$CS forms rapidly on the grains surface through hydrogenation of HCS:

\begin{eqnarray}
	\text{s-HCS} + \text{s-H} \to \text{s-H$_2$CS}
\end{eqnarray}

but is ultimately hydrogenated in either CH$_2$SH or CH$_3$S via:

\begin{eqnarray}
	\begin{split}
	\text{s-H$_2$CS} + \text{s-H}& \to &\text{s-CH$_2$SH}\\
	&\to& \text{s-CH$_3$S}
	\end{split}
\end{eqnarray}

In low density clouds however, H$_2$CS is formed much slower on grains surface. Hence at the end of the initial cloud runs, the differences in H$_2$CS abundances between a slow and a fast collapsing parent cloud can reach up to 2 orders of magnitude. These differences can afterwards vary during the collapse, since fast collapsing cells will reach high temperatures faster than slow collapsing cells. Indeed, once thermally desorbed in the gas phase at high temperatures, H$_2$CS abundance can endure significant variations \citep[see][]{Vidal18}, which could explain the range of abundances that corresponds to fast collapsing cells ($t_{\text{ff}} < 50$ kyrs) in figure \ref{fig_34}. For slow collapsing cells, H$_2$CS can continue to form on grains surface at low density before its thermal desorption. At that time however, the reactants of its main formation and destruction reactions at high temperature regime, such as atomic sulphur and CH$_3$, are scarce in the gas phase, which could explain the smaller range of abundances for slow collapsing cells ($t_{\text{ff}} > 50$ kyrs).

\subsection{NS}

\begin{figure}
        \begin{center}
	        \includegraphics[scale=0.45]{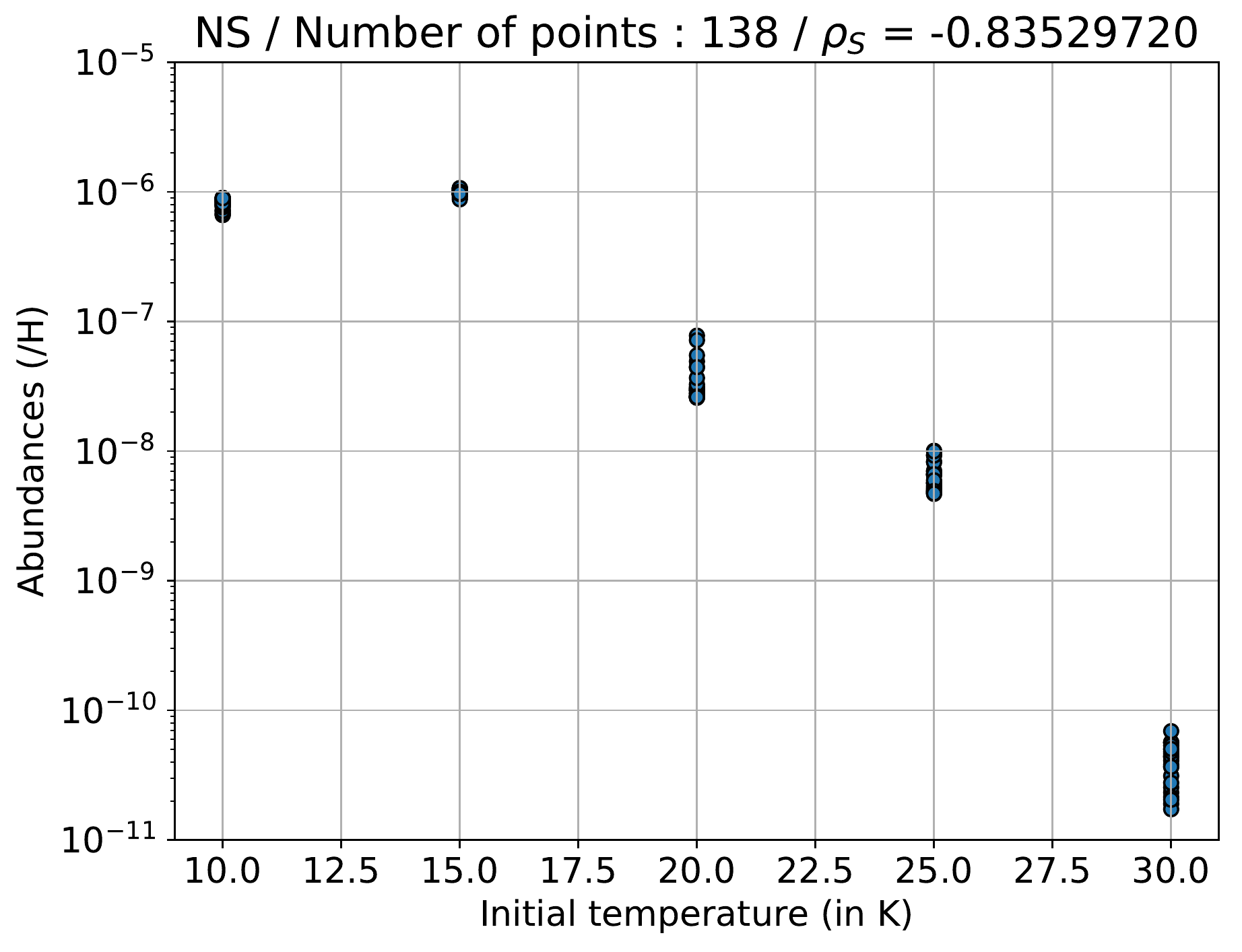}
                \caption{Distribution of the abundances of NS relative to H at the final time of the collapse as a function of the initial temperature.}
                \label{fig_35}
        \end{center}
\end{figure}

Figure \ref{fig_35} displays the abundance distribution in the HCR of mononitrogen monosulfide (NS) and its good anti-correlation with the initial temperature of the collapsing clouds with a Spearman's coefficient of -0.835. The corresponding Pearson's correlation coefficient is also low (-0.843), which implies a nearly linear relationship between NS abundance and the initial temperature. As such, this relationship defines NS as a precise tracer of the initial temperature of parent clouds. Hence, for an observed value of NS in a hot corino $[NS]_{\text{obs}}$, one could deduce the following:

\begin{itemize}
\item If $[NS]_{\text{obs}}$ is close to $10^{-6}$, $T_0 \leq 15$ K,\\
\item If $10^{-8} < [NS]_{\text{obs}} \leq 10^{-7}$, $T_0 \approx 20$ K,\\
\item If $10^{-9} < [NS]_{\text{obs}} \leq 10^{-8}$, $T_0 \approx 25$ K,\\
\item If $[NS]_{\text{obs}} < 10^{-10}$, $T_0 \geq 30$ K.
\end{itemize}

In the chemical model, NS is at first efficiently formed in the initial cloud on the grains surface from the main reservoir of sulphur HS \citep[see][]{Vidal17} via:

\begin{eqnarray}
	\text{s-N} + \text{s-HS} \to \text{s-NS} + \text{s-H}
\end{eqnarray}

This reaction is efficient in all the initial runs, regardless of the initial temperature, and the NS formed on the surfaces tends to sink in the grains bulk. However, it appears that the higher the temperature of the cloud, the more atomic oxygen can diffuse in the grains bulk. This has for main effect to destroy NS in the bulk via:

\begin{eqnarray}
	\text{b-O} + \text{b-NS} \to \text{b-NO} + \text{b-S}
\end{eqnarray}

This destruction of NS is therefore more efficient at high temperatures, which results in a difference between its abundance in initial clouds at 10 K and 30 K that can go up to four orders of magnitude. Moreover, once NS thermally desorbs in the hot gas, it is consumed by the following reaction:

\begin{eqnarray}
	\text{CH$_2$} + \text{NS} \to \text{H} + \text{HCNS}
\end{eqnarray}

Where the CH$_2$ comes from gas phase photodissociation by secondary UV photons of CH$_4$, which itself forms from the CH$_3$ produced by the destruction of thermally desorbed methanol (CH$_3$OH) and methyl formate (HCOOCH$_3$). This reaction is believed to be more efficient for cells that spend more time in a high temperature regime, which explains why those coming from parent clouds with $T_0 =30$ K have such a low NS abundances compared to those coming from parents clouds with  $T_0 =10$ K.\\
However, this result should be taken with caution since work is still ongoing to complete the NS chemical network, as discussed in \citet{Vidal17}.

\subsection{OCS}

\begin{figure}
        \begin{center}
	        \includegraphics[scale=0.45]{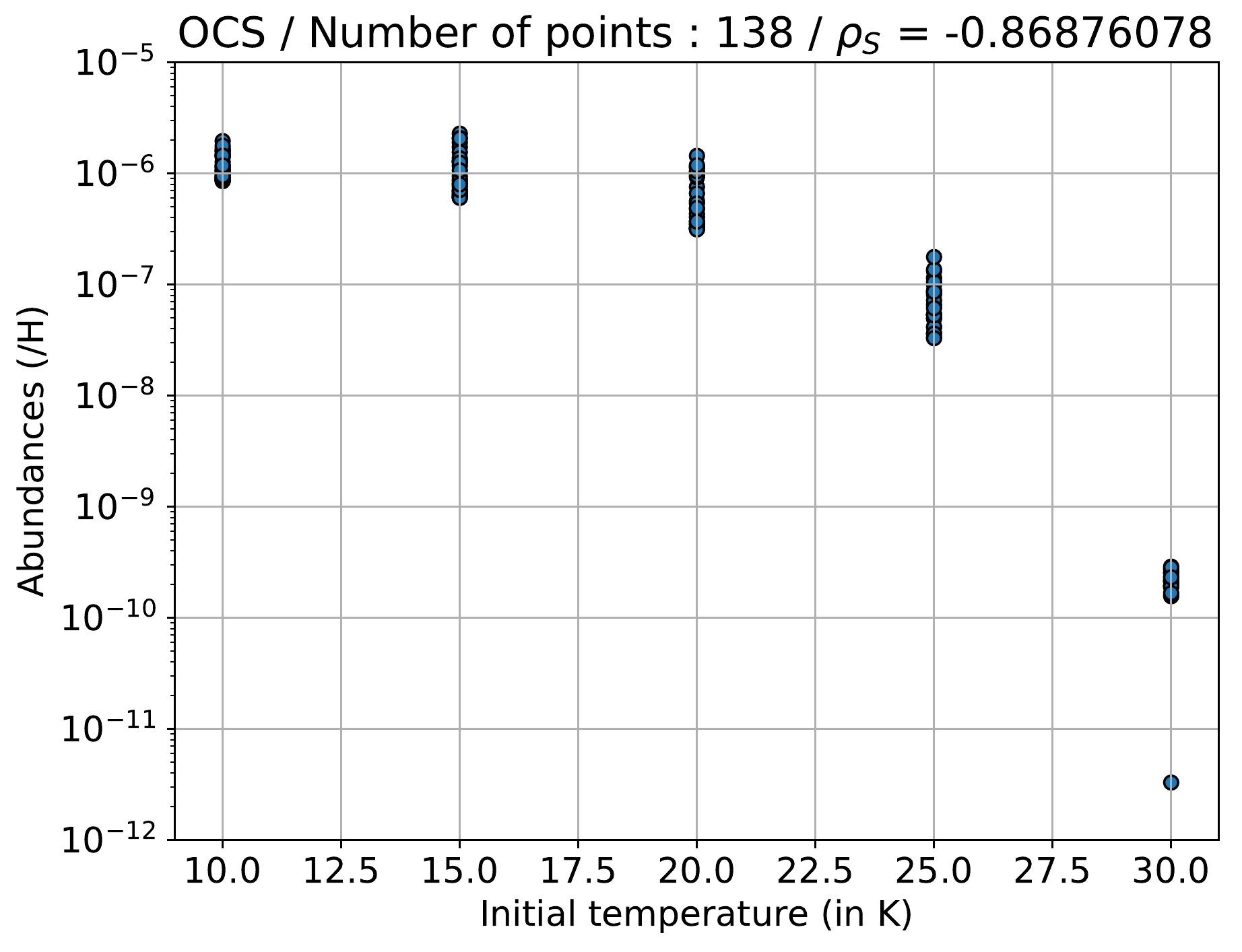}
                \caption{Distribution of the abundances of OCS relative to H at the final time of the collapse as a function of the initial temperatures.}
                \label{fig_36}
        \end{center}
\end{figure}

Figure \ref{fig_36} displays the abundances distribution in the HCR of carbonyl sulfide (OCS), which is similar to the NS one. Indeed, as for NS, a nearly-linear relationship exists between OCS abundances in the HCR and the initial temperatures of the collapsing clouds, with a Spearman's coefficient of -0.869. Hence, OCS appears to be a very good tracer of the initial temperature. For an observed value of OCS in a hot corino $[OCS]_{\text{obs}}$, one could deduce the following:

\begin{itemize}
\item If $[OCS]_{\text{obs}}$ is close to $10^{-6}$, $T_0 \leq 20$ K,\\
\item If $[OCS]_{\text{obs}}$ is close to $10^{-7}$, $T_0 \approx 25$ K,\\
\item If $[OCS]_{\text{obs}}<10^{-9}$, $T_0 \geq 30$ K.
\end{itemize}

It should be noted that this characterization is less precise than the one deduced from NS correlation, however we are much more confident in the completeness of the OCS chemical network than in the NS one, and therefore more confident about OCS correlation results.\\
From a chemical point of view, as for the previous species, the difference in OCS abundance as a function of the initial temperature can be explained with the chemistry of the initial cloud runs. Indeed, in all of them OCS is at first mainly produced in the gas phase via:

\begin{eqnarray}
	\text{CS} + \text{OH} \to \text{H} + \text{OCS}
\end{eqnarray}

It is also produced on the grains surface via:

\begin{eqnarray}
	\text{s-S} + \text{s-CO} \to \text{s-OCS}
	\label{ocs_surf}
\end{eqnarray}

However, the higher the temperature of the initial cloud is, the less OCS is formed on the grains surface. Indeed, at low temperature, reaction (\ref{ocs_surf}) stays efficient until the end of the parent cloud model because CO remains abundant on the grains surface. When the temperature rises however, the surface OH originated from the photodissociation by CR induced UV photons of water and methanol, which are both abundant on the grains surface, seems to efficiently consume the remaining CO through:

\begin{eqnarray}
	\begin{split}
	\text{s-OH} + \text{s-CO}& \to &\text{s-CO$_2$} + \text{s-H}\\
	&\to& \text{s-HOCO}
	\end{split}
\end{eqnarray}

This reaction is believed to become efficient at high temperatures because of the increase diffusivity of OH on grains surface. Indeed, this mechanism causes differences between OCS abundances in initial clouds at 10 K and 30 K that can go up to five orders of magnitude, which would explain the corresponding correlation plot in figure \ref{fig_36}. After thermal desorption, gas phase abundance of OCS should not vary more than one order of magnitude until the end of the collapse, as studied in \citet{Vidal18}.\\

\section{Comparison with observations} \label{compobs}

We now present the part of the study which focuses on the ER dataset, \textit{i.e.} the cells that do not exceed a temperature of 100 K at the final time of their respective models (see section \ref{regime}). Indeed, because the correlations between species abundances and IPPC in this region are not conclusive, we decided to exploit this part of the dataset in another way. Hence, we developed a simple method to efficiently constrain the IPPC from comparison with observations of Class 0 protostar envelopes. In the following, we first describe the method and the observational dataset, then continue on the presentation of a comprehensive exemple of the application of the method to the envelope of IRAS 16293-2422. Finally, we present the summary of the method applied to a sample of 12 sources.

\subsection{Presentation of the method} \label{method}

\begin{table*}
\caption{Summary of the observational dataset used for the method. $N_{\text{obs}}$ is the number of observed species used. }
	\begin{adjustbox}{center}
		\begin{tabular}{p{33mm} c p{43mm} c}
		\hline
		\hline
   		Sources & $N_{\text{obs}}$ & Observed species used & References \\
   		\hline
		\hline
		IRAS 16293-2422  &  20     &  CO, HCO$^+$, CN, HCN, HNC, HC$_3$N, CH$_3$CN, CCH, c-C$_3$H$_2$, CH$_3$CCH, H$_2$CO, CH$_3$OH, CS, SO, SO$_2$, OCS, HCS$^+$, H$_2$CS, SiO, N$_2$H$^+$      & \citet{Schoier02} \\
		\hline
		NGC1333-IRAS2A & 12 & CO, CS, SO, HCO$^+$, N$_2$H$^+$, HCN, HNC, CN, HC$_3$N, H$_2$CO, CH$_3$OH, CH$_3$CN & \citet{Jorgensen04,Jorgensen05}\\
		\hline
		NGC1333-IRAS4A, NGC1333-IRAS4B, LDN1448-C, VLA1623 & 11&CO, CS, SO, HCO$^+$, N$_2$H$^+$, HCN, HNC, CN, HC$_3$N, H$_2$CO, CH$_3$OH & \citet{Jorgensen04,Jorgensen05}\\
		\hline
		LDN1527, LDN483,\newline LDN723, LDN1157 & 10 & CO, CS, SO, HCO$^+$, N$_2$H$^+$, HCN, HNC, CN, HC$_3$N, H$_2$CO & \citet{Jorgensen04,Jorgensen05}\\
		\hline
		LDN1448-IRAS2 & 9 & CO, CS, SO, HCO$^+$, N$_2$H$^+$, HCN, HNC, CN, HC$_3$N & \citet{Jorgensen04} \\
		\hline
		LDN1551-IRAS5 & 7 &  CO, CS, SO, HCO$^+$, N$_2$H$^+$, HNC, HC$_3$N & \citet{Jorgensen04}\\
		\hline
		\hline
 		\end{tabular}
	\end{adjustbox}
  	\label{tab_33}
\end{table*}

The set of observational data used for the comparison with observations includes 12 Class 0 protostar envelopes toward which 7 to 20 species included in the chemical network had been observed. This set of observational data comprises the single -dish observations of \citet{Schoier02}, \citet{Jorgensen04}, and \citet{Jorgensen05} and is summarized in table \ref{tab_33}. Note that in the case of the observations of \citet{Schoier02}, we use the abundances derived for the outer part of the envelope thanks to a jump model (third column of table 7 in \citet{Schoier02}). In each of these papers, the ability of the model to reproduce the observational constraints is quantified using the $\chi^2$ statistics.

\begin{figure}
        \begin{center}
	        \includegraphics[scale=0.45]{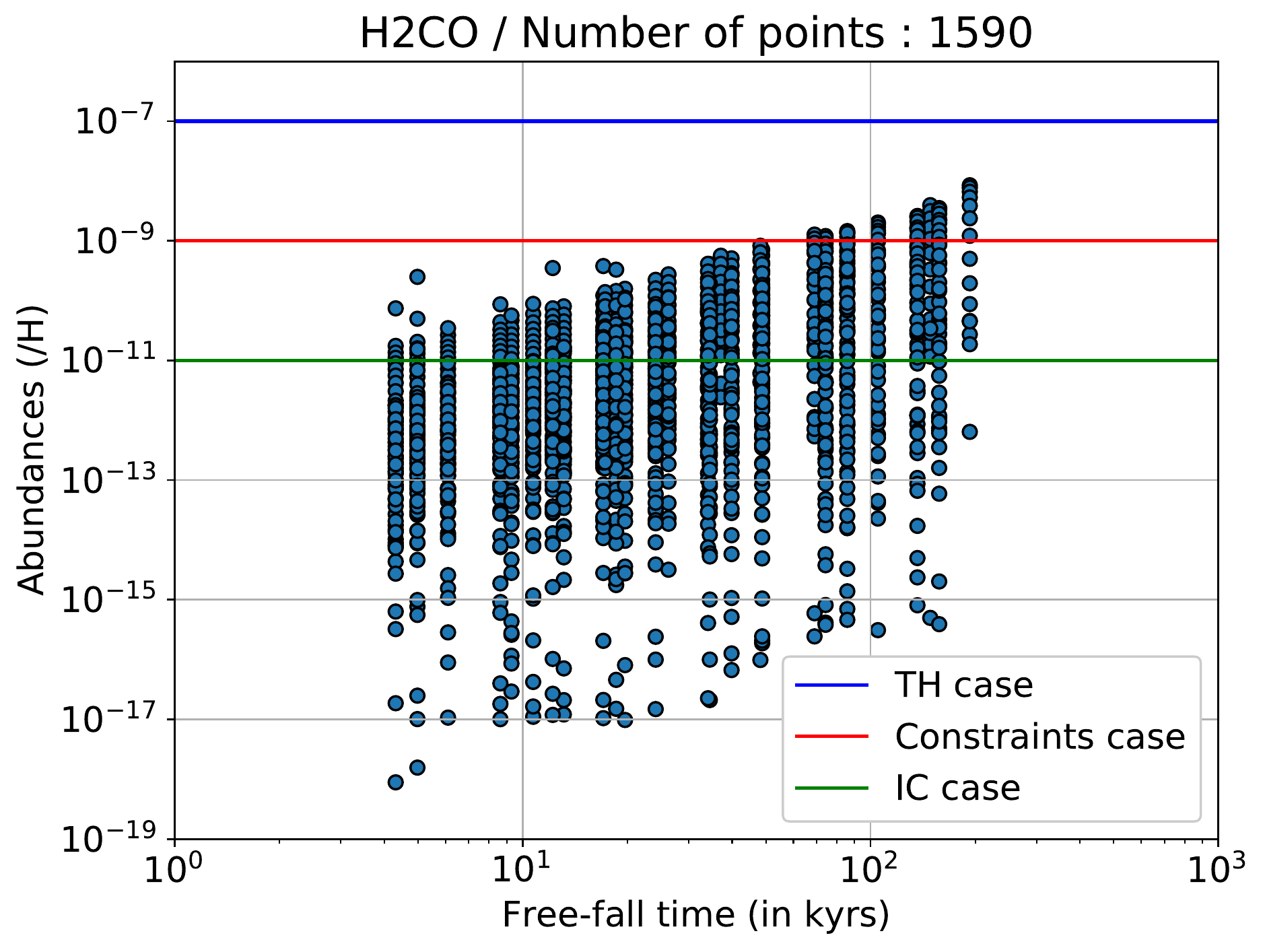}
                \caption{Example of the 3 possible cases of constraints that can be derived from the method with the distribution of the abundances of H$_2$CO relative to H at the final time of the collapse  as a function of the free-fall time. The blue line represents the TH case, the red one the case when constraints can be derived, and the green one the inconclusive case.}
                \label{fig_37}
        \end{center}
\end{figure}

Given the aforementioned dataset of observed species and their respective abundances in a Class 0 protostellar envelope, the principle of the method is to compare for each species its observed abundance to its corresponding abundance distributions in the ER for each of the IPPC considered in this study: $M_0$, $T_0$, $R_0$, $\rho_0$ and $t_{\text{ff}}$. As showed in figure \ref{fig_37}, the results of such comparison can then either be one of the following:

\begin{itemize}
\item The observed abundance is higher than all the abundances in the ER distribution, and no constraints can be found. We note this case TH (Too High),\\
\item The observed abundance intersects its respective ER distributions such as only a part of the studied IPPC range is compatible with the observations. Hence, the resulting constraint on the IPPC values will either be a single value, an interval, an upper limit or a lower limit. It should be noted that the latter are in fact also intervals since they are only defined within the studied IPPC range of values (see Reference Dataset column in table \ref{tab_31}).\\
\item The observed abundance intersects its respective ER distributions such as all the studied IPPC values are compatible, which means that no constraints on the IPPC can be found within its considered range of values. we note this case IC (InConclusive) in the following.
\end{itemize}

We do not characterize the case when the observed abundance is lower than all the abundances in the ER distribution. Indeed, for a species to be observable in the gas phase of the envelope, its abundance should be at least higher than 10$^{-12}$, and most of species ER distribution covers a range of abundances that nearly always goes well below this limit. Hence, for each species not in the TH case, the method returns an interval of possible values for each IPPC. The final possible intervals for the source are then obtained for each IPPC as the intersection of all the corresponding intervals obtained with each species.\\

Finally, because $\chi^2$ data on calculated abundances are not available for all species in the observational results used, we define a systematic procedure in order to ensure consistency while taking into account uncertainties on observed abundances. For each species we proceed as follows:

\begin{itemize}
\item If the corresponding $\chi^2$ value is available, we use an error margin of a factor of 3 on the observed abundance if $\chi^2 < 10$, and of a factor of 10 if $\chi^2 \geq 10$,\\
\item If the corresponding $\chi^2$ value is unavailable, we use the number of lines $N_L$ as a criteria for the error margin definition, and use an error margin of a factor of 3 if $N_L >1$, and of a factor of 10 otherwise.
\end{itemize}

\begin{figure}
        \begin{center}
	        \includegraphics[scale=0.45]{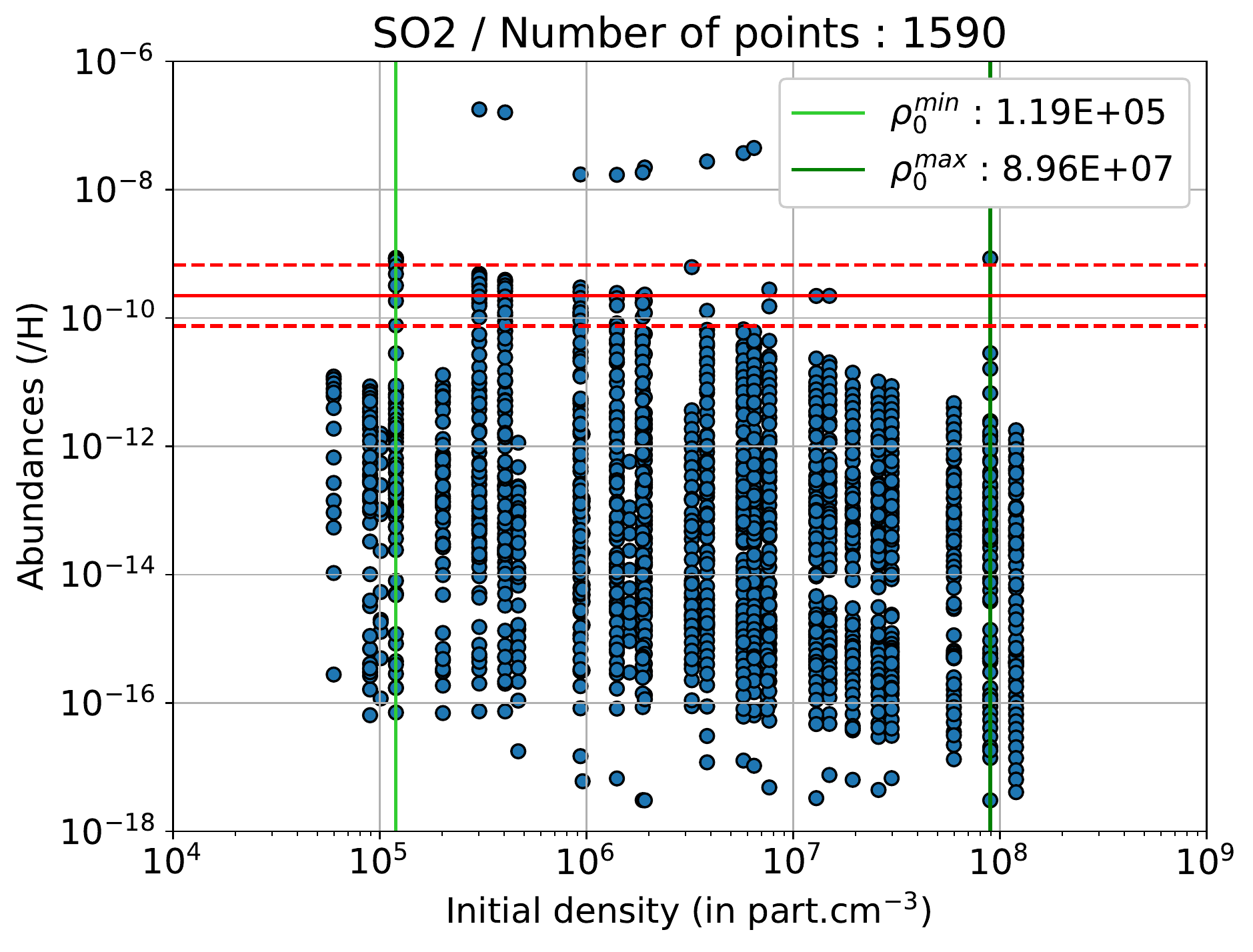}
                \caption{Comprehensive example of the application of the method on observed SO$_2$ in IRAS 16293-2422. The horizontal red line represents the observed abundance relative to H of $2.25\times10^{-10}$ and the dashed ones its uncertainty factor of 3. The light and dark green vertical lines respectively represent the minimal and maximal values of the initial density $\rho_0$ that are constrained by the observed abundance.}
                \label{fig_38}
        \end{center}
\end{figure}

In figure \ref{fig_38}, we display a comprehensive example of the application of the method to one species. we consider here SO$_2$ in IRAS 16293-2422, which has an observed abundance of $2.25\times10^{-10}$, and search for constraints on the initial density of the parent cloud. An error margin of factor 3 is used because the $\chi^2$ value is 0.5 (calculated with $N_L = 10$). The figure shows that the observed abundance can only be compatible with an initial density of the parent cloud between $1.20\times10^5$  and  $8.96\times10^7$ part.cm$^{-3}$. 

\subsection{Results on the envelope of IRAS 16293-2422}

\begin{table*}
\caption{Summary of the application our method to \citet{Schoier02}'s observations of IRAS 16293-2422. IC stands for inconclusive and ND means that no data were available. Upper and lower limits are only defined within their corresponding IPPC range of values (see table \ref{tab_31}). In red are contradictory results, and in blue the final constraints on the IPPC of the source. $e$ is the uncertainty factor.}
	\footnotesize\setlength{\tabcolsep}{2.5pt}
		\begin{tabular*}{\textwidth}{c @{\extracolsep{\fill}} cccccccc}
		\hline
		\hline
   		Species & $M_0$ ($M_\odot$) & $T_0$ (K) & $R_0$ (au) & $\rho_0$ (part.cm$^{-3}$) & $t_{\text{ff}}$ (kyrs) & $\chi^2$ & $N_L$ &	$e$ \\	
		\hline
		\hline
		CO	& IC & 15 & IC & $\geq 8.96E+04$ & $\leq157.52$ & 0.2 & 3 & 3\\
		\hline
		HCO$^+$ & $\geq 2$ & $\leq20$ & \textcolor{red}{$\geq20000$} & \textcolor{red}{$\leq$2.02E+05} & \textcolor{red}{$\geq105.01$}	 & 10.9 & 3 & 10\\
		\hline
		CN & IC & IC & IC & IC & IC & 1.2 & 4 & 3\\
		\hline
		HCN & IC & IC & IC & $\leq$ 8.96E+07 & $\geq4.98$ & 2.3 & 3 & 3\\
		\hline
		HNC & IC & IC & IC & $\leq$ 8.96E+07 & $\geq4.98$ & 0.2 & 3 & 3\\
		\hline
		HC$_3$N & IC & IC & IC & $\leq$ 2.99E+07 & $\geq8.63$ & 0.1 & 3 & 3\\
		\hline
		CH$_3$CN & $\geq 2$ & IC & IC & $\leq$ 8.96E+07 & $\geq4.98$ & 0.9 & 7 & 3\\
		\hline
		CCH & IC & IC & IC & IC & IC & 0.3 & 4 & 3\\
		\hline
		c-C$_3$H$_2$ & IC & IC & IC & IC & IC & 5.7 & 6 & 3\\
		\hline
		CH$_3$CCH & $\geq 1$ & IC & IC & [1.20E+05 ; 8.96E+07] & [4.98 ; 136.41] & 1.7 & 6 & 3\\
		\hline
		H$_2$CO & IC & IC & IC & $\leq$ 8.96E+07 & $\geq$ 4.98 & 1.9 & 7 & 	3\\
		\hline
		CH$_3$OH & IC & IC & IC & IC & IC & 1.2 & 23 & 3\\ 
		\hline
		CS & IC & IC & $\geq5000$ & $\leq$ 3.82E+06 & $\geq24.11$ & 0.5 & 3 & 3\\
		\hline
		SO & $\geq 1$ & $\leq25$ & IC & $\leq$ 8.96E+07 & $\geq4.98$ & 1.8 & 9 & 3\\
		\hline
		SO$_2$ & $\geq 1$ & IC & IC & [1.20E+05 ; 8.96E+07] & [4.98 ; 136.41] & 0.5 & 10 & 3\\
		\hline
		OCS & IC & $\leq25$ & IC & IC & IC & 1.5 & 7 & 3\\
		\hline
		HCS$^+$ & IC & IC & $\geq5000$ & $\leq$ 3.23E+06 & $\geq26.25$ & 7.0 & 2 & 3\\
		\hline
		H$_2$CS & IC & IC & IC & $\leq$ 5.97E+07 & $\geq6.10$ & 1.4 & 6,4 & 3\\
		\hline
		SiO & $\geq 1$ & IC & $\leq20000$ & [3.02E+05 ; 8.96E+07] & [4.98 ; 85.74] & 0.8 & 8 & 3\\
		\hline
		N$_2$H$^+$ & $\geq 1$ & IC & IC & IC & IC & ND & 1 & 10\\
		\hline
		\hline
		\textcolor{blue}{Constraints} & \textcolor{blue}{[2 ; 8]} & \textcolor{blue}{15} & \textcolor{blue}{[5000 ; 20000]} & \textcolor{blue}{[3.02E+05 ; 3.23E+06]} & \textcolor{blue}{[26.25 ; 85.74]} & & &\\
		\hline
		\hline
 		\end{tabular*}
  	\label{tab_34}
\end{table*}

Table \ref{tab_34} shows the results of the application of the method to \citet{Schoier02}'s observations of IRAS 16293-2422. This source is an ideal example because the high number of species considered allows to display all the possible types of constraint defined in section \ref{method} are found. For example, the constraint on $T_0$ given by CO is a single value (15 K) compatible with the upper limits given by HCO$^+$, SO and OCS, when the $\rho_0$ and $t_{\text{ff}}$ constraints given by CH$_3$CCH, SO$_2$, and SiO are intervals. However, as for the other sources, it appears that the main type of constraints is lower or upper limits.\\

The result, which is by far the most interesting, is that all species give a coherent set of constraints but one. In the present case it is HCO$^+$ that shows contradictory constraints on $R_0$, $\rho_0$ and $t_{\text{ff}}$. The $\chi^2$ value derived from the observations for this molecule is however the largest one in our sample, indicated an uncertain observed abundance. Additional studies of this molecule by \citet{Quenard18b}, using a 3D modeling of HCO$^+$ and its isotopologues emission in IRAS 16293-2422, showed that the contribution of the envelope of the HCO$^+$ emission is negligible compared to the contribution of the outflows. Such detailed analysis was however not done for other molecules for which this could be the case. The disagreement of our model with HCO$^+$ abundance may well reflect a higher cosmic-ray ionization rate as compared to the small values that we have used. In fact, \citet{Quenard18b}'s best model was for a ten times higher cosmic-ray ionization rate.\\

It appears that the method allows to successfully give constraints on the IPPC of the molecular cloud in which IRAS 16293-2422 has formed (in blue in table \ref{tab_34}). Indeed, the fact that 95\% (19/20) of the species-specific constraints agree with each other, which give confidence in the obtained results. Moreover, the constrained intervals for the envelope mass and radius are compatible with their respective values derived in \citet{Schoier02} of 5.4 M$_{\odot}$ and 6685 au. From an astrochemical point of view, these results, and in particular those regarding initial density and temperature, force to reconsider the value commonly used in chemical models to simulate the initial cloud in which IRAS 16293-2422 had formed. Indeed, to this time, the common set of values used were usually $\leq$ 10 K for the initial temperature and a few 10$^4$ part.cm$^{-3}$ for the initial density \citep[see for example][]{Majumdar17,Quenard18,Vidal18}. The present results suggest however that the values astrochemist should use are slightly higher, namely an initial density within the interval [3.02$\times10^5$ ; 3.23$\times10^6$] part.cm$^{-3}$ and an initial temperature of 15 K. This result is in agreement with those of \citet{Jaber17}, who inferred from cyanopolyynes observations and chemical modeling that IRAS 16293-2422 underwent a collapse with a $t_{\text{ff}} \leq 100$ kyrs, and those of \citet{Taquet18} who derived upper limits on the density and temperature of the parent dark cloud of IRAS 16293-2422 from deep ALMA search for O$_2$.

\subsection{Summary of the results on the source sample}

\subsubsection{Summary per source}

\begin{table*}
\caption{Summary of the constraints on the IPPC obtained with the method on the 12 studied sources. The percentage of agreement (noted PA) represents the number of species agreeing with the final constraints.}
	\begin{adjustbox}{center}
		\begin{tabular}{l c c c c c c c}
		\hline
		\hline
   		Sources & $M_0$ ($M_\odot$) & $T_0$ (K) & $R_0$ (au) & $\rho_0$ (part.cm$^{-3}$) & $t_{\text{ff}}$ (kyrs) & PA\\	
		\hline
		\hline
		IRAS 16293 & [2 ; 8] & 15 & [5000 ; 20000] & [3.02E+05 ; 3.23E+06] & [26.25 ; 85.74] &95.0\% \\
		\hline
		LDN1448-IRAS2 & [1 ; 8] & [10 ; 20] & IC & [8.96E+04 ; 2.58E+07] & [9.28 ; 157.52] & 88.9\% \\
		\hline
		LDN1448-C & [1 ; 8] & 15 & [5000 ; 30000] & [8.96E+04 ; 5.74E+06] & [19.69 ; 157.52] & 90.9\% \\
		\hline
		NGC1333-IRAS2A  & 4 & 10 & 30000 & 5.98E+04 & 192.92 & 91.7\% \\
		\hline
		NGC1333-IRAS4A & [2 ; 8] & 10 & [20000 ; 30000] & [5.98E+04 ; 2.01E+05] & [105.01 ; 192.92]& 90.9\% \\
		\hline
		NGC1333-IRAS4B & 2 &10 & 20000 & 1.01E+05 & 148.51 & 90.9\% \\
		\hline
		LDN1527 & [2 ; 8] & [10 ; 15] & [20000 ; 30000] & [8.96E+04 ; 1.20E+05] & [136.41 ; 157.52] & 100\% \\
		\hline
		VLA1623 & [1 ; 8] & 15 & [5000 ; 20000] & [3.02E+05 ; 7.65E+06] & [17.05 ; 85.74] & 90.9\% \\
		\hline
		LDN483 & [2 ; 4] & 10 & [20000 ; 30000] & [5.98E+04 ; 1.01E+05] & [148.51 ; 192.92] & 90.0\%\\
		\hline
		LDN723	 & 4 & 10 & 30000 & 5.98E+04 & 192.92& 90.0\% \\
		\hline
		LDN1157 & [2 ; 8] & [10 ; 20] & [20000 ; 30000] & [8.96E+04 ; 1.20E+05] & [136.41 ; 157.52] & 90.0\% \\
		\hline
		LDN1551-IRAS5 & [2 ; 4] & 10 & [20000 ; 30000] & [8.96E+04 ; 1.01E+05] & [148.51 ; 157.52] & 85.7\% \\
		\hline
		\hline
 		\end{tabular}
	\end{adjustbox}
  	\label{tab_36}
\end{table*}

Table \ref{tab_36} presents a summary of the IPPC constraints we find for each source listed in table \ref{tab_33} using the method described above (for a more comprehensive description of the results, see appendix \ref{consdetail}). As the table shows, the main result of this study is that the method allows to derive constraints on all the IPPC for all sources, except for $R_0$ in the case of LDN1448-IRAS2. The latter is due to the fact that for 7 of the 9 species considered for this source we applied an uncertainty factor of 10 because of the low quality of the corresponding observations (see table \ref{tab_D1} in appendix \ref{consdetail}), making it the least constrained source of the study.\\

\begin{figure}
        \begin{center}
	        \includegraphics[scale=0.32]{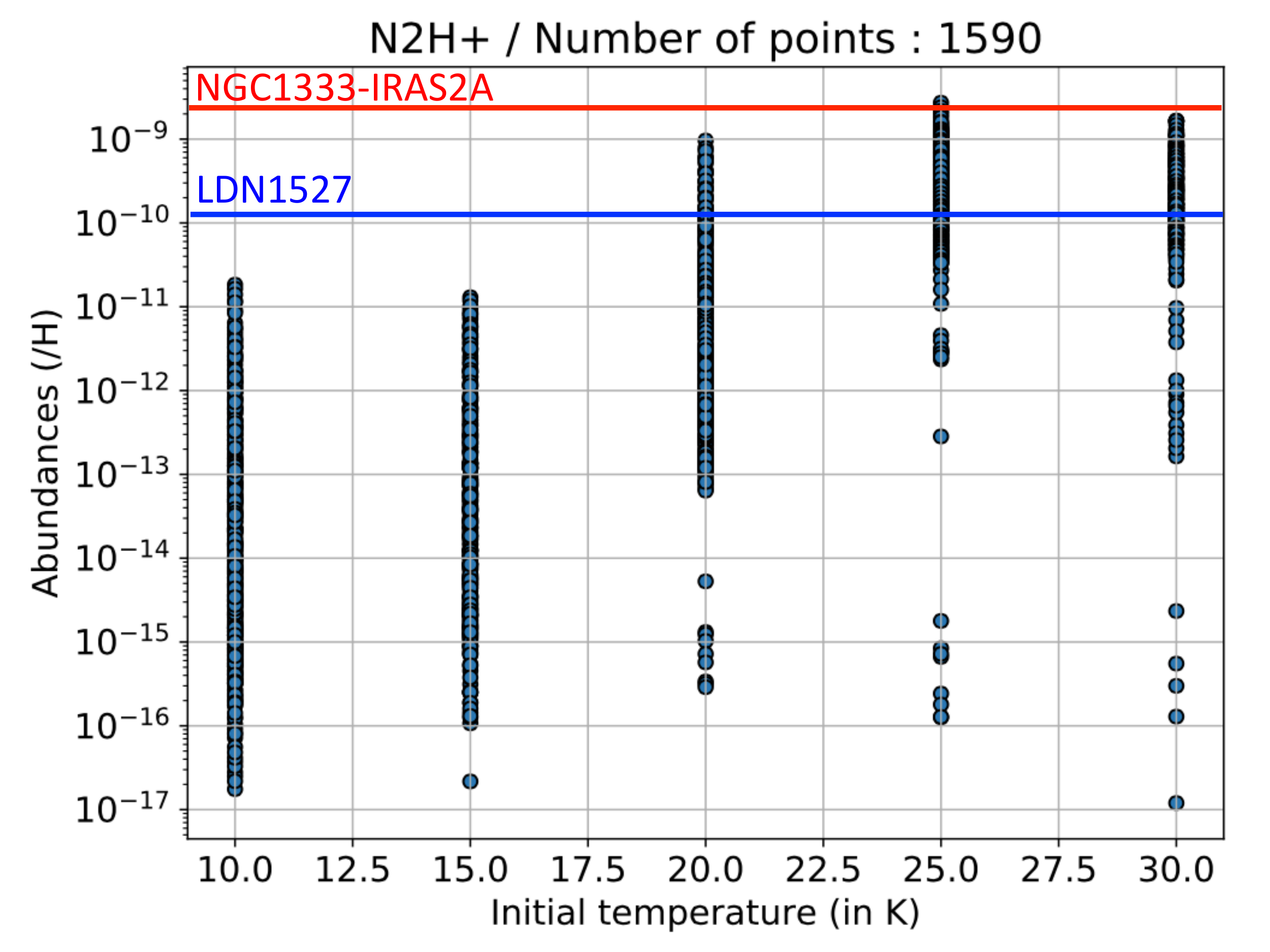}
                \caption{Abundances distribution of N$_2$H$^+$ relative to H in the ER as a function of $T_0$. As examples, the red and blue horizontal lines represent the observed abundances of N$_2$H$^+$ in NGC1333-IRAS2A and LDN1527, respectively.}
                \label{fig_39}
        \end{center}
\end{figure}

Another interesting result is that at most one species per source gives contradictory constraints except for LDN1527 for which the percentage of agreement, \textit{i.e.} the percentage of species giving results that agree with the final constraint, is 100\%. Hence, the percentage of agreement for the sources considered is always higher than 85\%. Furthermore, except in the case of IRAS 16293-2422 (see the previous section), the species that gives contradictory constraints for the concerned sources is always N$_2$H$^+$. Indeed, as can be seen in figure \ref{fig_39}, N$_2$H$^+$ abundances distribution in the ER as a function of $T_0$ is such that either its observed abundance is too high (above a few 10$^{-10}$) and corresponds to $T_0 \geq 20$ K (\textit{e.g.} for NGC1333-IRAS2A as displayed by the red horizontal line in figure \ref{fig_39}), either it is too low, which is the case for LDN1527 (see the horizontal blue line in figure \ref{fig_39}), and the result is inconclusive. In the first case, the resulting constraint on $T_0$ is for all concerned sources contradictory with at least one other species constraint. This issue appears in 10 of the 11 sources from \citet{Jorgensen04}. \citet{Lique15} found that the simulated line intensities of N$_2$H$^+$ in dark clouds increase significantly when using the newly determined collisional rate coefficients of this species with H$_2$ compared to the one calculated based on collision with He, the latter being used in \citet{Jorgensen04,Jorgensen05}. This study indicates that the derived N$_2$H$^+$ abundances might be overestimated. Another possibility, as for the underestimation of HCO$^+$ in IRAS16293-2422, would be that our adopted cosmic-ray ionization rate is too small. Using a larger $\zeta$ in our chemical model, would produce larger N$_2$H$^+$ abundances at low temperature. Therefore, we decided to ignore the constraint of N$_2$H$^+$ on $T_0$ in the study.\\


Table \ref{tab_36} also displays that for NGC1333-IRAS2A, NGC1333-IRAS4B and LDN723, all IPPC are constrained to single values. These single values constraints should be taken with caution, since this kind of results is model-dependent, \textit{i.e.} the direct consequences of the discretization of the parameters space in the collapse models database.\\

Another interesting feature of table \ref{tab_36} is that $M_0$ is the least constrained IPPC. Coupled with the results found on correlation in the HCR in section \ref{tracers} (see table \ref{tab_32}), this result hints the complexity of obtaining constraints on the initial mass of the parent cloud of forming star using chemistry. On the contrary, $T_0$ and $\rho_0$ (and consequently $t_{\text{ff}}$) appear to be easily constrainable with chemistry, which make sense since temperature and density variation throughout star formation are two of the main parameters affecting the chemistry. Finally, regarding $R_0$, it appears that we are able to derive relatively precise constraints on its values which could seem puzzling at first since the link between the chemistry of a source and the initial radius of its parent cloud is not easy to apprehend. However, equation (\ref{radiuscorr}), which links the $R_0$ to the central density $\rho_c$ explains the efficiency of the method on deriving constraints on $R_0$.\\

\subsubsection{Summary per IPPC}

\begin{figure*}
        \begin{center}
	        \includegraphics[scale=0.3]{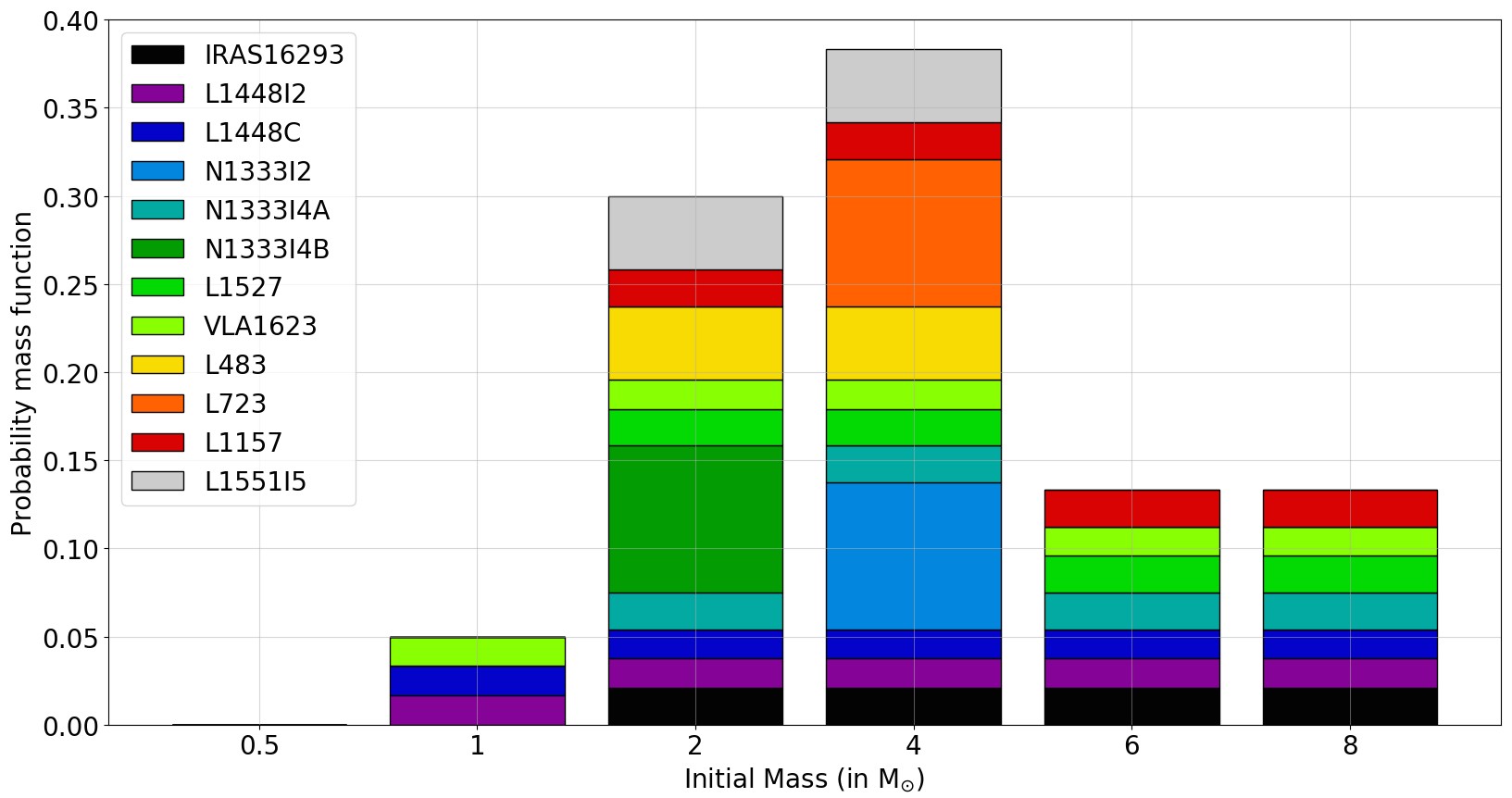}
                \caption{Probability mass function on the sample of 12 sources obtained for $M_0$.}
                \label{fig_311}
        \end{center}
\end{figure*}

In order to study the repartition of the source constraints for each IPPC, we plot for each of them their respective probability mass functions on the sample of sources. Hence the probability of finding a low-mass protostar with a particular value of the IPPC is calculated via:

\begin{eqnarray}
	P(y_i) = \frac{1}{N_s}\sum\limits_{s=1}^{N_s} \delta_{is}\frac{1}{N_y(s)}
\end{eqnarray}

where $Y = \{y_i\}_{i\in[1,N_y^{\text{tot}}]}$ is the distribution of possible values of the IPPC considered, $N_s = 12$ is the total number of sources considered, $\delta_{is}$ is the Dirac function assessing the match between the source $s$ and the IPPC value $y_i$, and $N_y(s)$ is the number of IPPC values constrained for the source $s$. The corresponding plots display in color the contribution of each of the sources to the mass function. Consequently, in the following, we express the results in terms of probabilities for the sake of clarity. However, it should be noted that these probabilities are defined for a relatively small sample of sources (12). Therefore, the probabilities presented hereafter are likely to change for a larger sample. \\

Figure \ref{fig_311} displays the results for $M_0$ which is, as stated previously, the least constrained of the IPPC considered. Nevertheless, it appears that tendencies can still be inferred from these constraints. First, the probability of finding sources with $M_0 < 1$ M$_{\odot}$ is null, and of only of 5\% for a parent cloud of mass $M_0 = 1$ M$_{\odot}$. Second, parent clouds of mass $M_0 = 2$ M$_{\odot}$ and $M_0 = 4$ M$_{\odot}$ corresponds to a total probability of 68\%. Finally, higher mass parent clouds also agree with a significant fraction of the sources, but with a much lower probability than for the previous values, namely 26\%. This repartition of the source constraints hints that low-mass stars are born mostly from clouds of masses higher than 1 M$_{\odot}$. Furthermore, it highlights that a large fraction of low-mass stars should form from clouds of masses within the interval [2 ; 4] M$_{\odot}$.\\

\begin{figure*}
        \begin{center}
	       \includegraphics[scale=0.3]{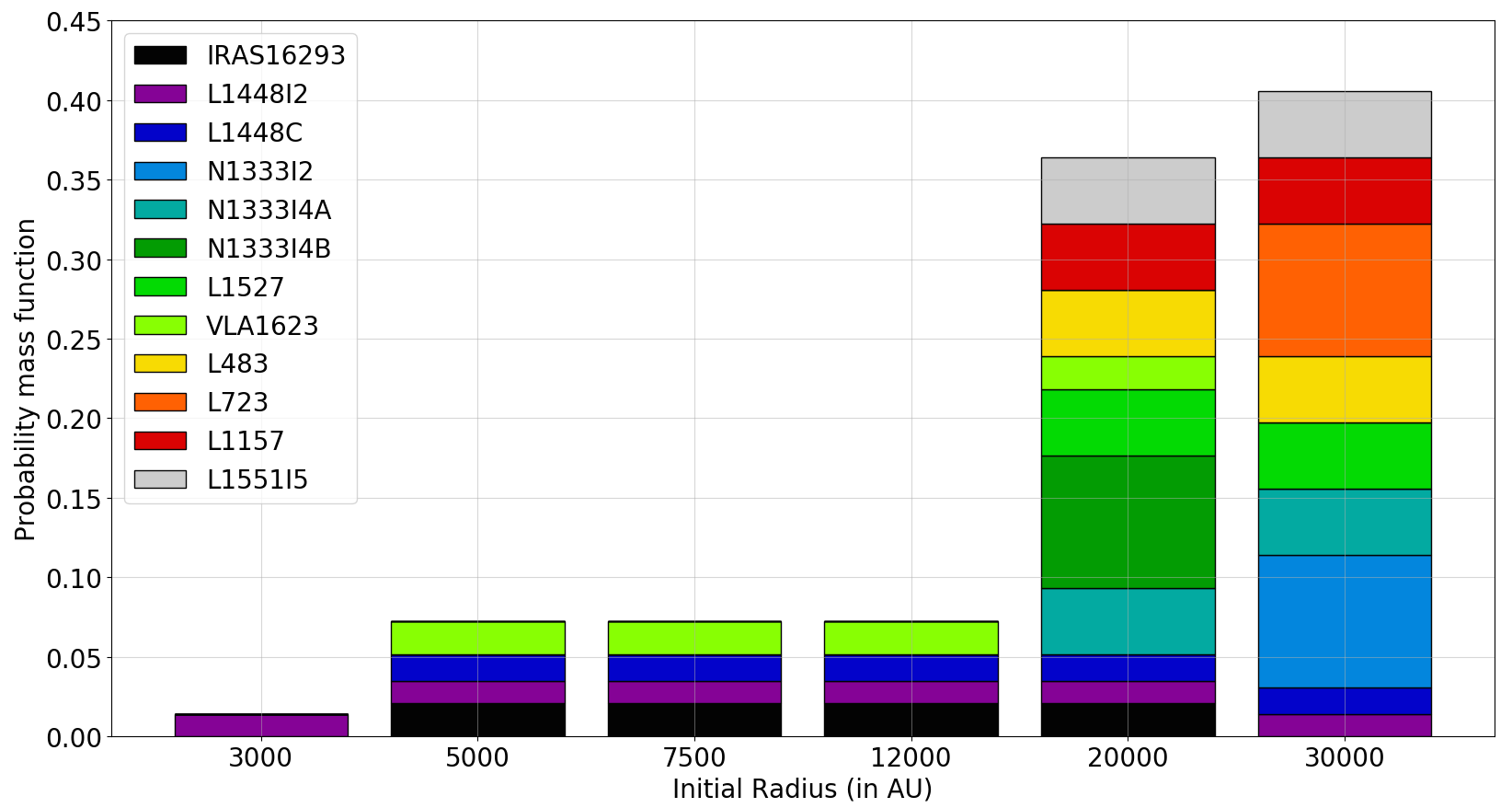}
                \caption{Probability mass function on the sample of 12 sources obtained for $R_0$.}
                \label{fig_312}
        \end{center}
\end{figure*}

Figure \ref{fig_312} shows the results obtained for $R_0$. As discussed previously, $R_0$ appears relatively well constrained by the method, since only two of its possible values agree with more than 4 of the sources with probabilities higher than 36\%. These values correspond to the largest parent clouds considered ($R_0 \geq 20000$ au), which suggests that low-mass stars have a total 77\% probability to be born from such large clouds. The probability for low-mass stars to form from clouds with radii within the interval [5000 ; 20000] au is rather low ($0.07\times3 = 21\%$) compared to the previous values. However, IRAS 16293-2422 and VLA 1623 are constrained within this interval, although with a large uncertainty on their initial radii, which hints that these sources represent particular cases of low-mass star formation. It should finally be noted that the probability to find sources corresponding to the smallest parent cloud (R$_0$ = 3000 au) is close to 0.\\

\begin{figure*}
        \begin{center}
	        \includegraphics[scale=0.3]{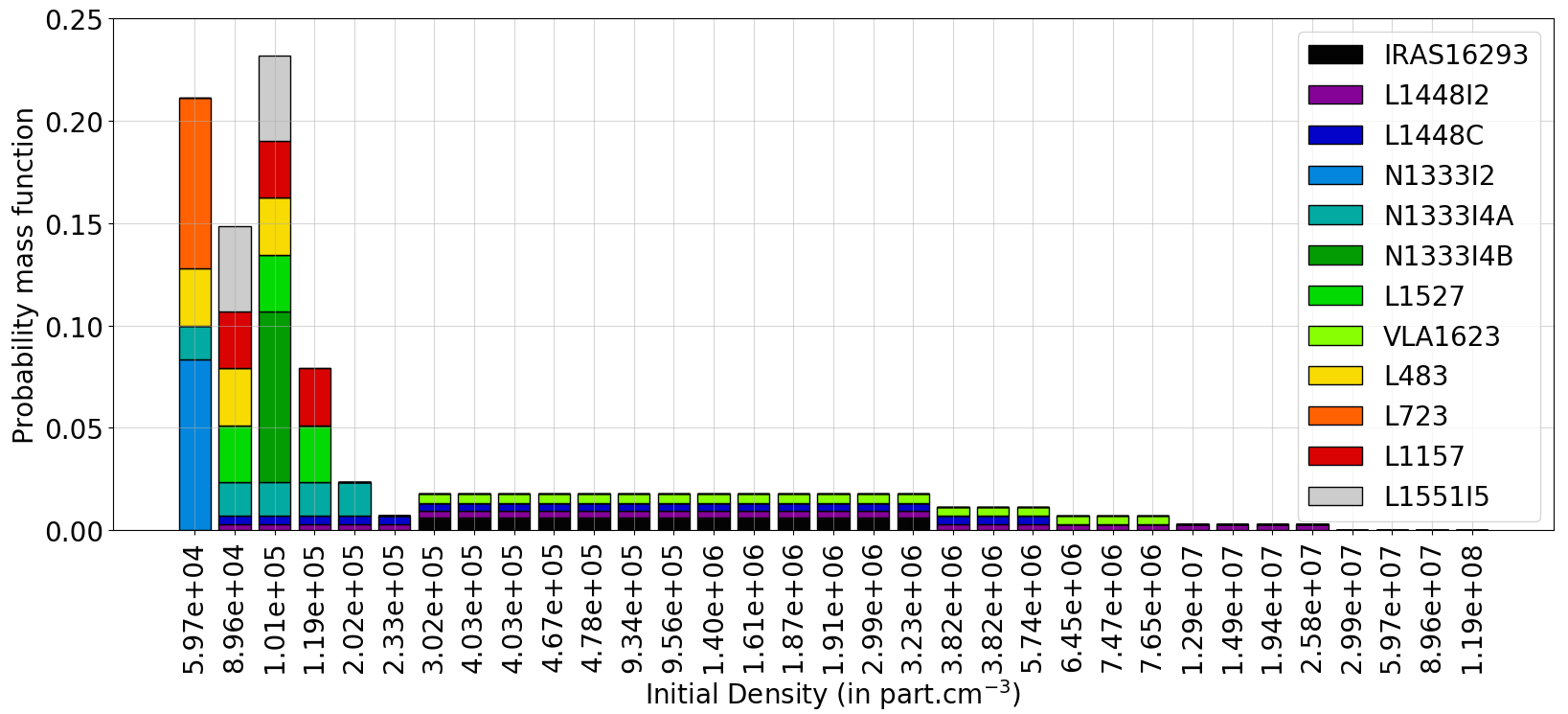}\\
	        \includegraphics[scale=0.3]{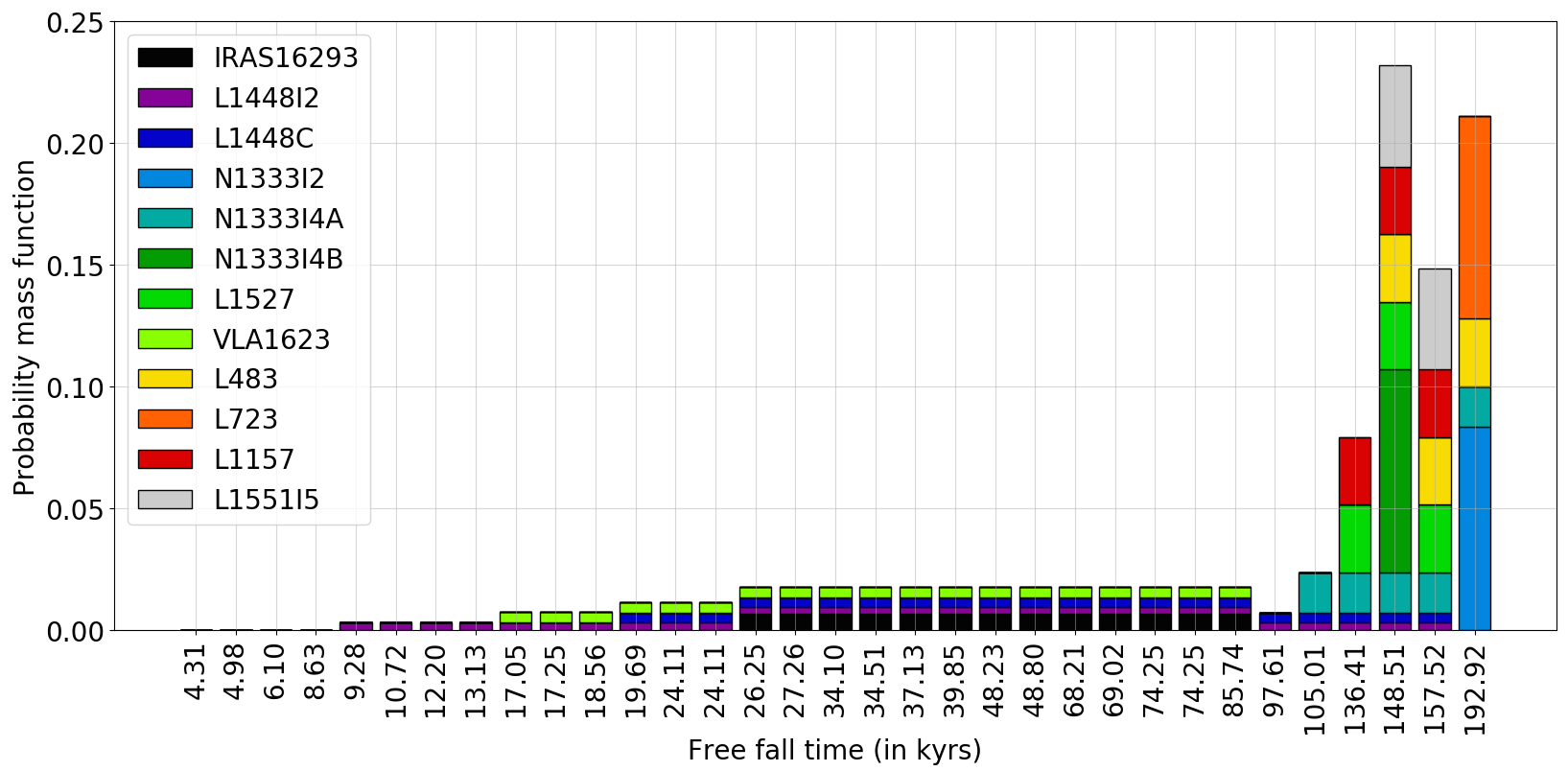}\\
                \caption{Probability mass function on the sample of 12 sources obtained for $\rho_0$ and $t_{\text{ff}}$.}
                \label{fig_313}
        \end{center}
\end{figure*}

The results regarding $\rho_0$ and $t_{\text{ff}}$ are displayed in figure \ref{fig_313}. As both these parameters are perfectly anti-correlated, the top figure appears as the mirror of the bottom one. One of the most interesting results regarding these IPPC is that clouds of density $> 10^7$ part.cm$^{-3}$ (respectively of $t_{\text{ff}} < 17$ kyrs) appear to be ruled out of the possible parent clouds of low-mass stars. Indeed, the probability of finding such sources is only of 0.8\%. Moreover, as the initial density increases (respectively $t_{\text{ff}}$ decreases), the probabilities decrease. Another interesting result is that 8 of the 12 sources are constrained within the interval [5.97$\times10^4$ ; 2.02$\times10^5$] part.cm$^{-3}$ (respectively $t_{\text{ff}} \in$ [105.01 ; 192.92] kyrs) indicating that approximately 70\% of low-mass stars are born in clouds with relatively low densities and high free-fall times. However, as found for R$_0$, VLA 1623 and IRAS 16293-2422 appear to represent a particular type of low-mass protostars since they are incompatible with these small values and would preferentially have formed in cloud with higher density between $3.02\times10^5$ and $3.23\times10^6$ part.cm$^{-3}$ (respectively $t_{\text{ff}} \in [26.25;85.74]$ kyrs), interval which corresponds to a rather low probability compared to the previous one, namely 21\%. This result hints that low-mass stars can form from clouds with densities (respectively free-fall times) within a rather large range of values.\\

\begin{figure*}
        \begin{center}
	        \includegraphics[scale=0.3]{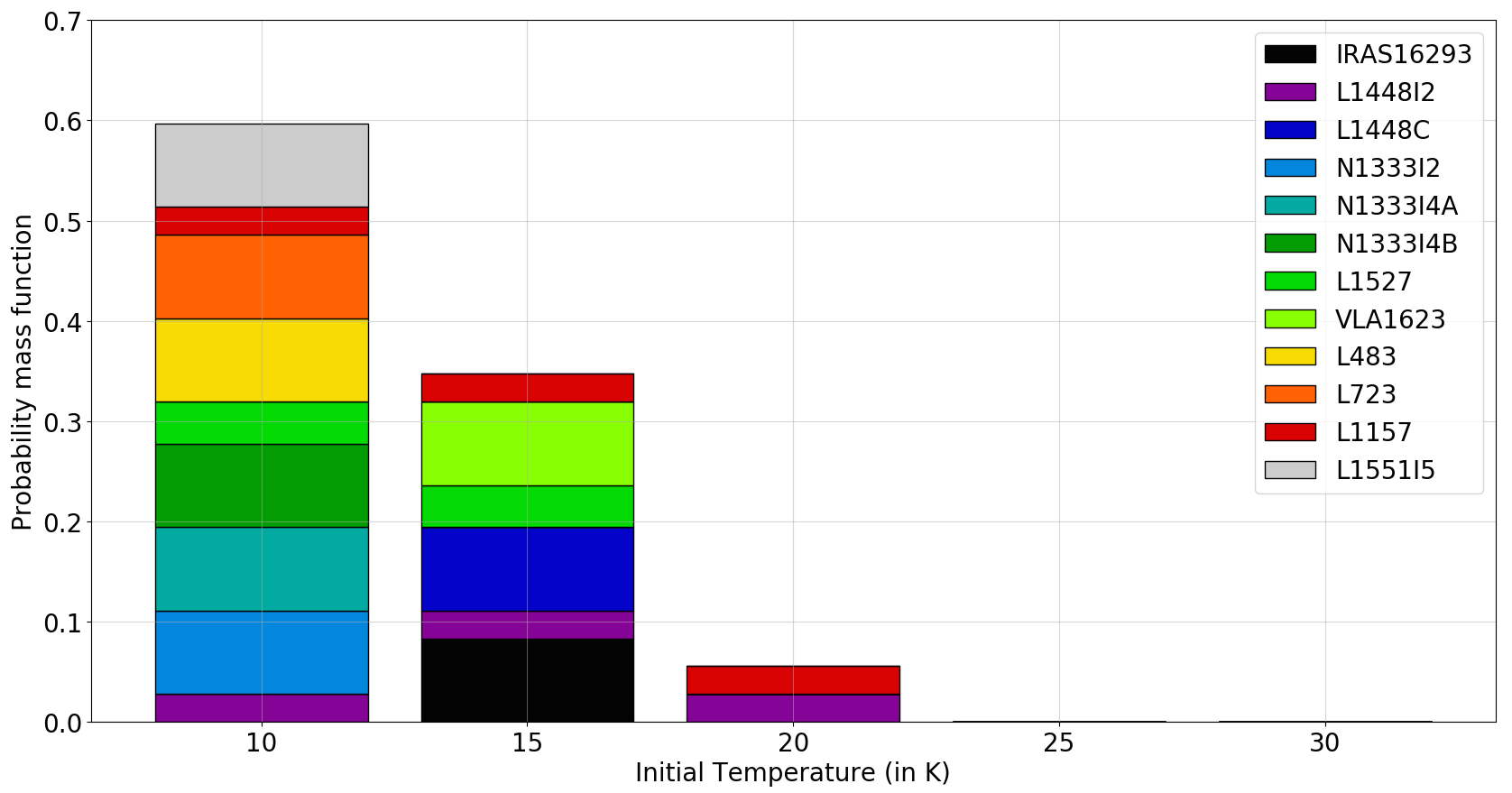}
                \caption{Probability mass function on the sample of 12 sources obtained for $T_0$.}
                \label{fig_314}
        \end{center}
\end{figure*}

Finally, figure \ref{fig_314} displays the results for $T_0$. This IPPC is the best constrained one, which results in a clearer repartition of the mass function. First, temperatures above 20 K appear to be ruled out of the possible values since none of the sources agreed with them. Second, all sources match with an initial temperature of either 10 or 15 K, corresponding to a total probability of 94\%. This shows a common trend among low-mass stars to form from a low temperature parent cloud, and particularly clouds with T$_0$ = 10 K. Finally, the two sources corresponding to $T_0 = 20$ K (LDN1448-IRAS2 and LDN1157) are both the least constrained sources regarding initial temperature. Since the resulting probability is only of 6\%, it appears that low-mass stars have a really small probability to form in clouds with temperatures as high as 20 K.\\

\section{Discussions}

In the following, we discuss the dependence of the results on the model used. Another aspect that we discuss below is the applicability of the method used in the ER to the HCR dataset.

\subsection{About the modeling bias} \label{bias}

Since we used the combination of two (physical and chemical) models, all the aforementioned results are model dependent. On the one hand, regarding the physical model, the completeness of the chemical abundances distribution over the range of the IPPC is limited by the discretization of the parameter space, especially for $M_0$ and $R_0$. Additionally, it appears for a majority of sources that the constrained lowest values of the initial densities and temperatures, as well as the highest values of the initial radii could be set by the boundaries of the parameter space. A more complete representation of these parameters could affect the results by modifying the abundance distribution of the species considered. Future similar study would therefore benefit of such a completion of the parameter space. \\ 

On the other hand, the temperature validity domain of the chemical network, especially its lower limit of 10 K forces limitations on the study. Indeed, as colder clouds can also theoretically form low-mass protostars, the chemical evolution of these clouds (available in the original dataset of \citet{Vaytet17}) could also fit the sources considered in the ER study and therefore enlarge the range of possible $T_0$ we found. To solve this issue, complementary studies on the modeling of the chemistry in such low-temperature regimes are still needed. Furthermore, as work remains to be conducted on the chemical network, especially for NS, the results on the NS correlation with $T_0$ are susceptible to change in the future. However the chemical network for CH$_3$CN, H$_2$CS and OCS have been completed recently \citep[][ Andron et al., submitted to MNRAS]{Vidal17},  which gives confidence in the correlation results presented for these species.\\

Finally, the most important modeling bias is due to the fact that we used 1D models. Indeed, compared to more comprehensive 2D or 3D collapse models, 1D models do not allow to access as much detailed information, especially regarding physical structure that form towards the center during collapse such as outflows or shocks at the centrifugal barrier \citep{Sakai14}. However 2D and 3D models require long computation times and imply large amount of output data, which would not have been compatible with the number of models and sources used in this study.

\subsection{About the applicability of the method on the HCR dataset} 

For the sake of completeness, we also tried to apply the method developed for the ER data on the joined TR and HCR data. Both sets were taken into account in order to ensure an optimal fit with the observed data. Indeed, the cells of the TR are located at radii smaller or of the same order of magnitude (tens of au, see figure \ref{fig_33}) as the spatial resolution of the observations of hot corinos we used. These observations are the data from \citet{Schoier02} on the hot corinos of the binary IRAS 16293-2422 obtained with their jump model, PILS data on the hot corino within IRAS 16293-2422B \citep{Jorgensen16,Coutens16,Lykke17,Drozdovskaya17} and \citet{Taquet15}'s COMs observations of the hot corinos NGC1333-IRAS2A and I4B. The number of species considered for each source was respectively 11, 9, 7 and 7. \\

Unfortunately for all sources, at least two species gave contradictory constraints on at least two IPPC, and we were unable to derive satisfactory constraints. This failure of the method on the TR+HCR dataset could be mostly explained by the following factors:

\begin{itemize}
\item An incompleteness of the gas phase high temperature network for some of the species considered, especially COMs. Moreover, a lack of understanding of high temperature chemical mechanism such as the formation of H$_2$ \citep[see][for a review on the subject]{Wakelam17} could also put uncertainties on the obtained chemical abundances in this region.\\
\item Uncertainties on the observed abundances, notably regarding beam dilution, since the sizes of the lobe obtained with the observations are bigger or of the same order of magnitude than the scale considered in the HCR. Moreover, at the considered scale, the envelopes and outflows could also alter the quality of the observations. Finally, it is also difficult to estimate the H$_2$ column density as the dust becomes optically thick.\\
\item The use of 1D models, since they do not allow to represent the complex physical structures that appear during the collapse at its center, such as outflows and the centrifugal barrier. Hence a lack of precision is expected regarding the modeling of the inner part of the collapsing cloud, and consequently the HCR.
\end{itemize} 

Nevertheless, the fact that the application of the method to the TR+HCR dataset was inconclusive does not contradict the results obtained for the ER one. In fact, we are confident in the completeness of the \textsc{Nautilus} chemical network for the species considered in the temperatures and densities regimes of the ER. Finally, the spatial scales considered in this region limit the uncertainties due to beam dilution.
\section{Conclusions}

In this article we have presented a large scale chemical study of Class 0 protostars formation, focusing on the constraint of five initial physical parameters of collapse (IPPC) of interest: $M_0$, $T_0$, $R_0$, $\rho_0$ and $t_{\text{ff}}$. The study was based on the association of a dataset of 110 1D physical models of collapsing clouds with the \textsc{Nautilus} chemical model. As the initial idea of the study was to use such a large dataset to infer statistical results, we first tried to find possible tracers of the IPPC by studying their respective correlations with each of the abundance distribution of the species present in the chemical network. In order to do so, we defined two physical regions in the protostellar envelope from the water abundance distribution: the Envelope Region (ER) corresponding to the the outer part of the protostellar envelope (T < 100 K) and the Hot Corino Region (HCR) corresponding to the hot corino (T>145 K). As the correlation obtained for the ER were mostly inconclusive, we focused the study on the correlation within the HCR. Despite the fact that we found no satisfactory correlations with $M_0$, the study shows for the four remaining IPPC  $T_0$, $R_0$, $\rho_0$ and $t_{\text{ff}}$ that numerous possible tracers were good candidates, such as  CH$_3$CN, H$_2$CS, NS and OCS. H$_2$CS was found to be a possible tracer for the free-fall time and the initial density, while the three remaining species are found to be possible tracers of the initial temperature of the parent cloud.\\

In order to work on the ER dataset, we also developed a simple method of comparison of the abundance distributions with the observations. From this method we were able to derive the IPPC for 12 Class 0 protostars from observations of 7 to 20 species towards their envelopes. The specific result obtained for each source could be of great interest for the future modeling of their chemistry. Looking at the results for each IPPC, we were also able to derive the following probabilities for the studied low-mass protostars:

\begin{itemize}
\item The probability for them to form from parent clouds with masses $\leq 1$ M$_{\odot}$ appears to be of only 5\%, while the favored interval [2 ; 4] M$_{\odot}$ reaches a probability of 68\%. The outermost interval [6 ; 8] M$_{\odot}$ corresponds to a total probability of 26\% and could therefore correspond to a non-negligible number of low-mass protostars,\\
\item The initial temperatures, which are the most constrained IPPC, are efficiently constrained between 10 and 15 K, with all sources agreeing with either or both values. Hence, the total probability for low-mass sources to form in clouds with such temperatures reaches 94\%. This result hints that, even if it is theoretically possible, low-mass stars would tend not to form from hotter parent clouds ($T_0 \geq 20$ K),\\
\item The initial radii are also well constrained with the method because of their correlation with initial densities. The result shows that low-mass stars form preferentially in vast parent clouds of $R_0 \geq 20000$ au, with a probability of 77\%. This limit is however very model dependent since the chosen value for $R_0$ in the original dataset are highly spaced when higher than 7500 au,\\
\item Finally, the most interesting result regarding the initial densities and the free-fall times is that low-mass stars appear to have a very low probability (0.8\%) to form in parent clouds of density higher than $10^7$ part.cm$^{-3}$ ($t_{\text{ff}} < 17$ kyrs). Moreover, 70\% of low-mass stars appear to be born in clouds with densities within the interval [5.97$\times10^4$ ; 2.02$\times10^5$] part.cm$^{-3}$ (respectively $t_{\text{ff}} \in$ [105.01 ; 192.92] kyrs). Finally the fact that the possible values obtained for all sources span between a few 10$^4$ and 10$^6$ part.cm$^{-3}$ (20 to 192 kyrs for the free-fall time) hints that low-mass stars can form in parent clouds with various possible densities (or free-fall times). 
\end{itemize}

These probabilities are based on a relatively small sample of 12 Class 0 low-mass protostars which could not be representative of all low-mass protostars. Therefore, the results of our study may be subjected to change by widening the sample of sources. Change in the result may also be expected from a wider parameter space as discussed in section \ref{bias}. Despite those two limitations, we are confident our results highlight robust tendencies regarding the initial physical parameters of formation of low mass protostars.\\

Another interesting results of this study is that two sources (IRAS 16293-2422 and VLA1623) have constraints on R$_0$, $\rho_0$ and $t_{\text{ff}}$ within intervals which are not the favored ones, hinting that they could be representatives of a specific type of protostars which forms from medium-sized clouds with higher densities (respectively smaller free-fall time) than the majority of low-mass protostars. In order to confirm this dual trend, a similar study with a higher number of sources should be conducted.\\ 

\section*{Acknowledgements}
This work has been founded by the European Research Council (Starting Grant 3DICE, grant agreement 336474). This work was supported by the Programme National "Physique et Chimie du Milieu Interstellaire" (PCMI) of CNRS/INSU with INC/INP co-funded by CEA and CNES.




\bibliographystyle{mnras}
\bibliography{bibliography} 



\appendix

\section{Constraints on the IPPC for each sources} \label{consdetail}

In this appendix, we display the detailed results of the application of the comparison to observations described in section \ref{method} to all the sources considered in table \ref{tab_32} with the exception of IRAS 16293-2422, which results are displayed in table \ref{tab_34}. Hence, the following tables are the same as table \ref{tab_34} but for respectively LDN1448-IRAS2, LDN1448-C, NGC1333-IRAS2A, NGC1333-IRAS4A, NGC1333-IRAS4B, LDN1527, VLA1623, LDN483, LDN723, LDN1157, and LDN1551-IRAS5. In these table, IC stands for InConclusive and ND means that no data were available. Upper and lower limits are only defined within their corresponding IPPC range of values (see table \ref{tab_31}). In red are contradictory results, and in blue the final constraints on the IPPC of the source. $e$ is the uncertainty factor.

\begin{table*}
\caption{Summary of the application of our method to \citet{Jorgensen04}'s observations of LDN1448-IRAS2.}
	\footnotesize\setlength{\tabcolsep}{2.5pt}
		\begin{tabular*}{\textwidth}{c @{\extracolsep{\fill}} cccccccc}
		\hline
		\hline
   		Species & $M_0$ ($M_\odot$) & $T_0$ (K) & $R_0$ (au) & $\rho_0$ (part.cm$^{-3}$) & $t_{\text{ff}}$ (kyrs) & $\chi^2$ & $N_L$ &	$e$ \\	
		\hline
		\hline
		CO	& IC & $\leq$ 20 & IC & IC & IC & ND & 4 & 3\\
		\hline
		CS	& IC & IC & IC & IC & IC & ND & 3 & 10\\
		\hline
		SO & IC & IC & IC & $\leq$ 8.96E+07 & $\geq$ 4.98 & ND & 1 & 10\\
		\hline
		HCO$^+$ & IC & IC & IC & $\leq$2.58E+07 & $\geq$ 9.28 & ND & 1 & 10\\
		\hline
		N$_2$H$^+$ & $\geq$1 & \textcolor{red}{$\geq$ 20} & IC & $\geq$ 8.96E+04 & $\leq$ 157.52 & ND & 1 & 10\\
		\hline
		HCN & IC & IC & IC & $\leq$ 8.96E+07 & $\geq$ 4.98 & ND & 1 & 10\\
		\hline
		HNC & IC & IC & IC & IC & IC & ND & 2 & 3\\
		\hline
		CN & IC & IC & IC & $\leq$ 5.97E+07 & $\geq$ 6.10 & 0.13 & 4 & 3\\
		\hline
		HC$_3$N	 & IC	 & IC	 & IC & $\leq$ 5.97E+07 & $\geq$ 6.10 & ND & 1 & 10\\ 
		\hline
		\hline
		\textcolor{blue}{Constraints} & \textcolor{blue}{[1 ; 8]} & \textcolor{blue}{[10 ; 20]} & \textcolor{blue}{[3000 ; 30000]} & \textcolor{blue}{[8.96E+04 ; 2.58E+07]} & \textcolor{blue}{[9.28 ; 157.52]} & & &\\
		\hline
		\hline
 		\end{tabular*}
  	\label{tab_D1}
\end{table*}

\begin{table*}
\caption{Summary of the application of our method to \citet{Jorgensen04,Jorgensen05}'s observations of LDN1448-C.}
	\footnotesize\setlength{\tabcolsep}{2.5pt}
		\begin{tabular*}{\textwidth}{c @{\extracolsep{\fill}} cccccccc}
		\hline
		\hline
   		Species & $M_0$ ($M_\odot$) & $T_0$ (K) & $R_0$ (au) & $\rho_0$ (part.cm$^{-3}$) & $t_{\text{ff}}$ (kyrs) & $\chi^2$ & $N_L$ &	$e$ \\	
		\hline
		\hline
		CO	& IC & 15 & IC & $\geq$ 8.96E+04 & $\leq$ 157.52 & 2.1 & 4 & 3\\
		\hline
		CS	& IC & IC & $\geq$ 5000 & $\leq$ 5.74E+06 & $\geq$ 19.69 & 2.6 & 2 & 3\\
		\hline
		SO & IC & IC & IC & $\leq$ 8.96E+07 & $\geq$ 4.98 & ND & 1 & 10\\
		\hline
		HCO$^+$ & IC & IC & IC & IC & IC & 0.21 & 4 & 3\\
		\hline
		N$_2$H$^+$ & IC & \textcolor{red}{$\geq$ 20} & IC & $\geq$ 8.96E+04 & $\leq$ 157.52 & ND & 1 & 10\\
		\hline
		HCN & $\geq$ 1 & IC & IC & $\leq$ 8.96E+07 & $\geq$ 4.98 & 2.5 & 3 & 3\\
		\hline
		HNC & IC & IC & IC & $\leq$ 8.96E+07 & $\geq$ 4.98 & ND & 2 & 3\\
		\hline
		CN & IC & IC & IC & $\leq$ 1.94E+07 & $\geq$ 10.72 & 2.4 & 4 & 3\\
		\hline
		HC$_3$N	 & IC	 & IC	 & IC & $\leq$ 5.97E+07 & $\geq$ 6.10 & ND & 1 & 10\\ 
		\hline
		H$_2$CO & IC & IC & $\geq$ 5000 & $\leq$ 7.65E+06 & $\geq$ 17.05 & 1.2 & 3 & 3\\
		\hline
		CH$_3$OH & IC & IC & IC & IC & IC & ND & 1 & 10\\
		\hline
		\hline
		\textcolor{blue}{Constraints} & \textcolor{blue}{[1 ; 8]} & \textcolor{blue}{15} & \textcolor{blue}{[5000 ; 30000]} & \textcolor{blue}{[8.96E+04 ; 5.74E+06]} & \textcolor{blue}{[19.69 ; 157.52]} & & &\\
		\hline
		\hline
 		\end{tabular*}
  	\label{tab_D2}
\end{table*}

\begin{table*}
\caption{Summary of the application of our method to \citet{Jorgensen04,Jorgensen05}'s observations of NGC1333-IRAS2A.}
	\footnotesize\setlength{\tabcolsep}{2.5pt}
		\begin{tabular*}{\textwidth}{c @{\extracolsep{\fill}} cccccccc}
		\hline
		\hline
   		Species & $M_0$ ($M_\odot$) & $T_0$ (K) & $R_0$ (au) & $\rho_0$ (part.cm$^{-3}$) & $t_{\text{ff}}$ (kyrs) & $\chi^2$ & $N_L$ &	$e$ \\	
		\hline
		\hline
		CO	& IC & $\leq$ 15 & IC & IC & IC & 3.9 & 4 & 3\\
		\hline
		CS	& IC & IC & $\geq$ 5000 & $\leq$ 3.82E+06 & $\geq$ 24.11 & 0.91 & 2 & 3\\
		\hline
		SO & $\geq$ 1 & $\leq$ 25 & IC & $\leq$ 8.96E+07 & $\geq$ 4.98 & 1.3 & 2 & 3\\
		\hline
		HCO$^+$ & 4 & 10 & 30000 & 5.98E+04 & 192.92 & 0.59 & 4 & 3\\
		\hline
		N$_2$H$^+$ & $\geq$ 1 & \textcolor{red}{$\geq$ 20} & IC & \textcolor{red}{$\geq$ 8.96E+04} & \textcolor{red}{$\leq$ 157.52} & ND & 1 & 10\\
		\hline
		HCN & IC & IC & IC & $\leq$ 8.96E+07 & $\geq$ 4.98 & 7.1 & 3 & 3\\
		\hline
		HNC & IC & IC & IC & $\leq$ 8.96E+07 & $\geq$ 4.98 & ND & 2 & 3\\
		\hline
		CN & IC & IC & IC & $\leq$ 5.97E+07 & $\geq$ 6.10 & 2.3 & 4 & 3\\
		\hline
		HC$_3$N	 & IC	 & IC	 & IC & $\leq$ 2.98E+07 & $\geq$ 8.63 & ND & 1 & 10\\ 
		\hline
		H$_2$CO & IC & IC & IC & $\leq$ 8.96E+07 & $\geq$ 4.98 & 0.63 & 3 & 3\\
		\hline
		CH$_3$OH & IC & IC & IC & IC & IC & 20.4 & 8 & 10\\
		\hline
		CH$_3$CN & $\geq$ 2 & IC & IC & $\leq$ 8.96E+07 & $\geq$ 4.98 & 8.5 & 4 & 3\\
		\hline
		\hline
		\textcolor{blue}{Constraints} & \textcolor{blue}{4} & \textcolor{blue}{10} & \textcolor{blue}{30000} & \textcolor{blue}{5.98E+04} & \textcolor{blue}{192.92} & & &\\
		\hline
		\hline
 		\end{tabular*}
  	\label{tab_D3}
\end{table*}

\begin{table*}
\caption{Summary of the application of our method to \citet{Jorgensen04,Jorgensen05}'s observations of NGC1333-IRAS4A.}
	\footnotesize\setlength{\tabcolsep}{2.5pt}
		\begin{tabular*}{\textwidth}{c @{\extracolsep{\fill}} cccccccc}
		\hline
		\hline
   		Species & $M_0$ ($M_\odot$) & $T_0$ (K) & $R_0$ (au) & $\rho_0$ (part.cm$^{-3}$) & $t_{\text{ff}}$ (kyrs) & $\chi^2$ & $N_L$ &	$e$ \\	
		\hline
		\hline
		CO	& IC & $\leq$ 20 & IC & IC & IC & 0.7 & 4 & 3\\
		\hline
		CS	& IC & IC & IC & $\leq$ 5.97E+07 & $\geq$ 6.10 & 0.019 & 2 & 3\\
		\hline
		SO & $\geq$ 1 & $\leq$ 25 & IC & $\leq$ 8.96E+07 & $\geq$ 4.98 & 0.66 & 2 & 3\\
		\hline
		HCO$^+$ & $\geq$ 2 & $\leq$ 20 & $\geq$ 20000 & $\leq$ 2.01E+05 & $\geq$ 105.01 & 1.2 & 3 & 3\\
		\hline
		N$_2$H$^+$ &  $\geq$ 1 & \textcolor{red}{$\geq$ 20} & IC & $\geq$ 8.96E+04 & $\leq$ 157.52 & ND & 1 & 10\\
		\hline
		HCN & IC & IC & IC & $\leq$ 8.96E+07 & $\geq$ 4.98 & 18 & 3 & 10\\
		\hline
		HNC & IC & IC & IC & IC & IC & ND & 2 & 3\\
		\hline
		CN & IC & IC & IC & IC & IC & 1.4 & 4 & 3\\
		\hline
		HC$_3$N	 & IC	 & IC	 & IC & $\leq$ 8.96E+07 & $\geq$ 4.98 & ND & 1 & 10\\ 
		\hline
		H$_2$CO & IC & IC & IC & $\leq$ 8.96E+07 & $\geq$ 4.98 & 0.59 & 3 & 3\\
		\hline
		CH$_3$OH & $\geq$ 2 & 10 & $\geq$ 5000 & $\leq$ 2.58E+07 & $\geq$ 9.28 & 4.5 & 7 & 3\\
		\hline
		\hline
		\textcolor{blue}{Constraints} & \textcolor{blue}{[2 ; 8]} & \textcolor{blue}{10} & \textcolor{blue}{[20000 ; 30000]} & \textcolor{blue}{[5.98E+04 ; 2.01E+05]} & \textcolor{blue}{[105.01 ; 192.92]} & & &\\
		\hline
		\hline
 		\end{tabular*}
  	\label{tab_D4}
\end{table*}

\begin{table*}
\caption{Summary of the application of our method to \citet{Jorgensen04,Jorgensen05}'s observations of NGC1333-IRAS4B.}
	\footnotesize\setlength{\tabcolsep}{2.5pt}
		\begin{tabular*}{\textwidth}{c @{\extracolsep{\fill}} cccccccc}
		\hline
		\hline
   		Species & $M_0$ ($M_\odot$) & $T_0$ (K) & $R_0$ (au) & $\rho_0$ (part.cm$^{-3}$) & $t_{\text{ff}}$ (kyrs) & $\chi^2$ & $N_L$ &	$e$ \\	
		\hline
		\hline
		CO	& IC & $\leq$ 20 & IC & IC & IC & 0.3 & 4 & 3\\
		\hline
		CS	& IC & IC & IC & $\leq$ 5.97E+07 & $\geq$ 6.10 & 1.1 & 2 & 3\\
		\hline
		SO & $\geq$ 1 & $\leq$ 25 & IC & $\leq$ 8.96E+07 & $\geq$ 4.98 & 0.82 & 2 & 3\\
		\hline
		HCO$^+$ & $\geq$ 2 & $\leq$ 20 & $\geq$ 20000 & $\leq$ 1.20E+05 & $\geq$ 136.41 & 2.7 & 2 & 3\\
		\hline
		N$_2$H$^+$ &  $\geq$ 1 & \textcolor{red}{$\geq$ 20} & IC & $\geq$ 8.96E+04 & $\leq$ 157.52 & ND & 1 & 10\\
		\hline
		HCN & IC & IC & IC & $\leq$ 8.96E+07 & $\geq$ 4.98 & 1.9 & 3 & 3\\
		\hline
		HNC & IC & IC & IC & $\leq$ 8.96E+07 & $\geq$ 4.98 & ND & 2 & 3\\
		\hline
		CN & IC & IC & IC & $\leq$ 8.96E+07 & $\geq$ 4.98 & 1.5 & 4 & 3\\
		\hline
		HC$_3$N	 & IC	 & IC	 & IC & $\leq$ 8.96E+07 & $\geq$ 4.98 & ND & 1 & 10\\ 
		\hline
		H$_2$CO &  $\geq$ 1 & IC &  $\geq$ 7500 & $\leq$ 9.56E+05 & $\geq$ 48.23 & 6.2 & 3 & 3\\
		\hline
		CH$_3$OH & 2 & 10 &  20000 & 1.01E+05 & 148.51 & 11.4 & 7 & 10\\
		\hline
		\hline
		\textcolor{blue}{Constraints} & \textcolor{blue}{2} & \textcolor{blue}{10} & \textcolor{blue}{20000} & \textcolor{blue}{1.01E+05} & \textcolor{blue}{148.51} & & &\\
		\hline
		\hline
 		\end{tabular*}
  	\label{tab_D5}
\end{table*}

\begin{table*}
\caption{Summary of the application of our method to \citet{Jorgensen04,Jorgensen05}'s observations of LDN1527.}
	\footnotesize\setlength{\tabcolsep}{2.5pt}
		\begin{tabular*}{\textwidth}{c @{\extracolsep{\fill}} cccccccc}
		\hline
		\hline
   		Species & $M_0$ ($M_\odot$) & $T_0$ (K) & $R_0$ (au) & $\rho_0$ (part.cm$^{-3}$) & $t_{\text{ff}}$ (kyrs) & $\chi^2$ & $N_L$ &	$e$ \\	
		\hline
		\hline
		CO	& IC & $\leq$ 15 & IC & $\geq$ 8.96E+04 & $\leq$ 157.52 & 1.0 & 4 & 3\\
		\hline
		CS	& IC & IC & IC & $\leq$ 8.96E+07 & $\geq$ 4.98 & 2.9 & 4 & 3\\
		\hline
		SO & IC & IC & IC & $\leq$ 8.96E+07 & $\geq$ 4.98 & 1.4 & 2 & 3\\
		\hline
		HCO$^+$ & $\geq$ 2 & $\leq$ 20 & $\geq$ 20000 & $\leq$ 1.20E+05 & $\geq$ 136.41 & 0.41 & 5 & 3\\
		\hline
		N$_2$H$^+$ &  $\geq$ 1 & IC & IC & IC & IC & ND & 1 & 10\\
		\hline
		HCN & IC & IC & IC & $\leq$ 8.96E+07 & $\geq$ 4.98 & ND & 1 & 10\\
		\hline
		HNC & IC & IC & IC & $\leq$ 8.96E+07 & $\geq$ 4.98 & ND & 2 & 3\\
		\hline
		CN & IC & IC & IC & $\leq$ 1.94E+07 & $\geq$ 10.72 & 0.93 & 4 & 3\\
		\hline
		HC$_3$N	 & $\geq$ 1 & IC & IC & $\leq$ 2.99E+07 & $\geq$ 8.62 & ND & 1 & 10\\ 
		\hline
		H$_2$CO & IC & IC & $\geq$ 5000 & $\leq$ 7.65E+06 & $\geq$ 17.05 & 7.7 & 3 & 3\\
		\hline
		\hline
		\textcolor{blue}{Constraints} & \textcolor{blue}{[2 ; 8]} & \textcolor{blue}{[10 ; 15]} & \textcolor{blue}{[20000 ; 30000]} & \textcolor{blue}{[8.96E+04 ; 1.20E+05]} & \textcolor{blue}{[136.41;157.52]} & & &\\
		\hline
		\hline
 		\end{tabular*}
  	\label{tab_D6}
\end{table*}

\begin{table*}
\caption{Summary of the application of our method to \citet{Jorgensen04,Jorgensen05}'s observations of VLA1623.}
	\footnotesize\setlength{\tabcolsep}{2.5pt}
		\begin{tabular*}{\textwidth}{c @{\extracolsep{\fill}} cccccccc}
		\hline
		\hline
   		Species & $M_0$ ($M_\odot$) & $T_0$ (K) & $R_0$ (au) & $\rho_0$ (part.cm$^{-3}$) & $t_{\text{ff}}$ (kyrs) & $\chi^2$ & $N_L$ &	$e$ \\	
		\hline
		\hline
		CO	& IC & 15 & IC & $\geq$ 8.96E+04 & $\leq$ 157.52 & 0.01 & 4 & 3\\
		\hline
		CS	& IC & IC & $\geq$ 5000 & $\leq$ 3.82E+06 & $\geq$ 24.11 & 2.6 & 2 & 3\\
		\hline
		SO & $\geq$ 1 & $\leq$ 25 & $\leq$ 20000 &  [3.02E+05 ; 8.96E+07] & [4.98 ; 85.74] & 1.6 & 3 & 3\\
		\hline
		HCO$^+$ & IC & IC & IC & IC & IC & 2.0 & 3 & 3\\
		\hline
		N$_2$H$^+$ &  $\geq$ 1 & \textcolor{red}{$\geq$ 20} & IC & $\geq$ 8.96E+04 & $\leq$ 157.52 & ND & 1 & 10\\
		\hline
		HCN & IC & IC & IC & $\leq$ 8.96E+07 & $\geq$ 4.98 & ND & 1 & 10\\
		\hline
		HNC & IC & IC & IC & $\leq$ 8.96E+07 & $\geq$ 4.98 & ND & 2 & 3\\
		\hline
		CN & IC & IC & IC & $\leq$ 2.98E+07 & $\geq$ 8.62 & 2.0 & 4 & 3\\
		\hline
		HC$_3$N	 & IC	 & IC	 & IC &$\leq$ 2.98E+07 & $\geq$ 8.62 & ND & 1 & 10\\ 
		\hline
		H$_2$CO & IC & IC & $\geq$ 5000 & $\leq$ 7.65E+06 & $\geq$ 17.05 & 0.061 & 3 & 3\\
		\hline
		CH$_3$OH & IC & IC &  IC & IC & IC & ND & 1 & 10\\
		\hline
		\hline
		\textcolor{blue}{Constraints} & \textcolor{blue}{[1 ; 8]} & \textcolor{blue}{15} & \textcolor{blue}{[5000 ; 20000]} & \textcolor{blue}{[3.02E+05 ; 7.65E+06]} & \textcolor{blue}{[17.05 ; 85.74]} & & &\\
		\hline
		\hline
 		\end{tabular*}
  	\label{tab_D7}
\end{table*}

\begin{table*}
\caption{Summary of the application of our method to \citet{Jorgensen04,Jorgensen05}'s observations of LDN483.}
	\footnotesize\setlength{\tabcolsep}{2.5pt}
		\begin{tabular*}{\textwidth}{c @{\extracolsep{\fill}} cccccccc}
		\hline
		\hline
   		Species & $M_0$ ($M_\odot$) & $T_0$ (K) & $R_0$ (au) & $\rho_0$ (part.cm$^{-3}$) & $t_{\text{ff}}$ (kyrs) & $\chi^2$ & $N_L$ &	$e$ \\	
		\hline
		\hline
		CO	& IC & $\leq$ 20 & IC & IC & IC & 1.1 & 4 & 3\\
		\hline
		CS	& IC & IC & IC & $\leq$ 8.96E+07 & $\geq$ 4.98 & 1.7 & 2 & 3\\
		\hline
		SO  & IC & IC & IC & $\leq$ 8.96E+07 & $\geq$ 4.98 & ND & 1 & 10\\
		\hline
		HCO$^+$ & [2 ; 4] & 10 & $\geq$ 20000 & $\leq$ 1.01E+05 & $\geq$ 148.51 & 0.63 & 4 & 3\\
		\hline
		N$_2$H$^+$ &  $\geq$ 1 & \textcolor{red}{$\geq$ 20} & IC & $\geq$ 8.96E+04 & $\leq$ 157.52 & ND & 1 & 10\\
		\hline
		HCN & IC & IC & IC & $\leq$ 8.96E+07 & $\geq$ 4.98 & 1.0 & 3 & 3\\
		\hline
		HNC & IC & IC & IC & $\leq$ 8.96E+07 & $\geq$ 4.98 & ND & 2 & 3\\
		\hline
		CN & IC & IC & IC & $\leq$ 2.99E+07 & $\geq$ 8.63 & 2.3 & 4 & 3\\
		\hline
		HC$_3$N	 & IC	 & IC	 & IC & $\leq$ 5.97E+07 & $\geq$ 6.10 & ND & 1 & 10\\ 
		\hline
		H$_2$CO & IC & IC & $\geq$ 5000 & $\leq$ 1.61E+06 & $\geq$ 37.13 & ND & 1 & 10\\
		\hline
		\hline
		\textcolor{blue}{Constraints} & \textcolor{blue}{[2 ; 4]} & \textcolor{blue}{10} & \textcolor{blue}{[20000 ; 30000]} & \textcolor{blue}{[5.98E+04 ; 1.01E+05]} & \textcolor{blue}{[148.51 ; 192.92]} & & &\\
		\hline
		\hline
 		\end{tabular*}
  	\label{tab_D8}
\end{table*}

\begin{table*}
\caption{Summary of the application of our method to \citet{Jorgensen04,Jorgensen05}'s observations of LDN723.}
	\footnotesize\setlength{\tabcolsep}{2.5pt}
		\begin{tabular*}{\textwidth}{c @{\extracolsep{\fill}} cccccccc}
		\hline
		\hline
   		Species & $M_0$ ($M_\odot$) & $T_0$ (K) & $R_0$ (au) & $\rho_0$ (part.cm$^{-3}$) & $t_{\text{ff}}$ (kyrs) & $\chi^2$ & $N_L$ &	$e$ \\	
		\hline
		\hline
		CO	& IC & $\leq$ 20 & IC & IC & IC & 2.5 & 4 & 3\\
		\hline
		CS	& IC & IC & IC & $\leq$ 8.96E+07 & $\geq$ 4.98 & 13 & 2 & 10\\
		\hline
		SO & $\geq$ 1 & IC & IC & $\leq$ 8.96E+07 & $\geq$ 4.98 & ND & 1 & 10\\
		\hline
		HCO$^+$ & 4 & 10 &  30000 & 5.98E+04 & 192.92 & 0.61 & 3 & 3\\
		\hline
		N$_2$H$^+$ &  $\geq$ 1 & \textcolor{red}{$\geq$ 20} & IC & $\geq$ 8.96E+04 & $\leq$ 157.52 & ND & 1 & 10\\
		\hline
		HCN & IC & IC & IC & $\leq$ 8.96E+07 & $\geq$ 4.98 & ND & 1 & 10\\
		\hline
		HNC & IC & IC & IC & $\leq$ 8.96E+07 & $\geq$ 4.98 & ND & 2 & 3\\
		\hline
		CN & IC & IC & IC & $\leq$ 2.99E+07 & $\geq$ 8.63 & 1.8 & 4 & 3\\
		\hline
		HC$_3$N	 &$\geq$ 1 & IC & IC &$\leq$ 2.99E+07 & $\geq$ 8.63 & ND & 1 & 10\\ 
		\hline
		H$_2$CO & IC & IC & IC & $\leq$ 8.96E+07 & $\geq$ 4.98 & ND & 1 & 10\\
		\hline
		\hline
		\textcolor{blue}{Constraints} & \textcolor{blue}{4} & \textcolor{blue}{10} & \textcolor{blue}{30000} & \textcolor{blue}{5.98E+04} & \textcolor{blue}{192.92} & & &\\
		\hline
		\hline
 		\end{tabular*}
  	\label{tab_D9}
\end{table*}

\begin{table*}
\caption{Summary of the application of our method to \citet{Jorgensen04,Jorgensen05}'s observations of LDN1157.}
	\footnotesize\setlength{\tabcolsep}{2.5pt}
		\begin{tabular*}{\textwidth}{c @{\extracolsep{\fill}} cccccccc}
		\hline
		\hline
   		Species & $M_0$ ($M_\odot$) & $T_0$ (K) & $R_0$ (au) & $\rho_0$ (part.cm$^{-3}$) & $t_{\text{ff}}$ (kyrs) & $\chi^2$ & $N_L$ &	$e$ \\	
		\hline
		\hline
		CO	& IC & $\leq$ 20 & IC & IC & IC & 3.3 & 4 & 3\\
		\hline
		CS	& IC & IC & IC & IC & IC & ND & 1 & 10\\
		\hline
		SO &  $\geq$ 1 & IC & IC & $\leq$ 8.96E+07 & $\geq$ 4.98 & 0.25 & 3 & 3\\
		\hline
		HCO$^+$ & $\geq$ 2 & $\leq$ 20 & $\geq$ 20000 & $\leq$ 1.20E+05 & $\geq$ 136.41 & 0.29 & 2 & 3\\
		\hline
		N$_2$H$^+$ &  $\geq$ 1 & \textcolor{red}{$\geq$ 20} & IC & $\geq$ 8.96E+04 & $\leq$ 157.52 & ND & 1 & 10\\
		\hline
		HCN & IC & IC & IC & IC & IC & ND & 1 & 10\\
		\hline
		HNC & IC & IC & IC & IC & IC & ND & 2 & 3\\
		\hline
		CN & IC & IC & IC & IC & IC & 1.1 & 4 & 3\\
		\hline
		HC$_3$N	 & IC	 & IC	 & IC &$\leq$ 8.96E+07 & $\geq$ 4.98 & ND & 1 & 10\\ 
		\hline
		H$_2$CO & IC & IC & IC & IC & IC & 3.9 & 3 & 3\\
		\hline
		\hline
		\textcolor{blue}{Constraints} & \textcolor{blue}{[2 ; 8]} & \textcolor{blue}{[10 ; 20]} & \textcolor{blue}{[20000 ; 30000]} & \textcolor{blue}{[8.96E+04 ; 1.20E+05]} & \textcolor{blue}{[136.41 ; 157.52]} & & &\\
		\hline
		\hline
 		\end{tabular*}
  	\label{tab_D10}
\end{table*}

\begin{table*}
\caption{Summary of the application of our method to \citet{Jorgensen04}'s observations of LDN1551-IRAS5.}
	\footnotesize\setlength{\tabcolsep}{2.5pt}
		\begin{tabular*}{\textwidth}{c @{\extracolsep{\fill}} cccccccc}
		\hline
		\hline
   		Species & $M_0$ ($M_\odot$) & $T_0$ (K) & $R_0$ (au) & $\rho_0$ (part.cm$^{-3}$) & $t_{\text{ff}}$ (kyrs) & $\chi^2$ & $N_L$ &	$e$ \\	
		\hline
		\hline
		CO	& IC & $\leq$ 15 & IC & IC & IC & 1.3 & 4 & 3\\
		\hline
		CS	& IC & IC & IC & $\leq$ 8.96E+07 & $\geq$ 4.98 & 0.64 & 4 & 3\\
		\hline
		SO &  $\geq$ 1 & IC & IC & $\leq$ 8.96E+07 & $\geq$ 4.98 & 0.95 & 2 & 3\\
		\hline
		HCO$^+$ & [2 ; 4] & 10 & $\geq$ 20000 & $\leq$ 1.01E+05 & $\geq$ 148.51 & 1.3 & 2 & 3\\
		\hline
		N$_2$H$^+$ &  $\geq$ 1 & \textcolor{red}{$\geq$ 20} & IC & $\geq$ 8.96E+04 & $\leq$ 157.52 & ND & 1 & 10\\
		\hline
		HNC & IC & IC & IC & $\leq$ 8.96E+07 & $\geq$ 4.98 & ND & 2 & 3\\
		\hline
		HC$_3$N	 & IC	 & IC	 & IC &$\leq$ 5.97E+07 & $\geq$ 6.10 & ND & 1 & 10\\ 
		\hline
		\hline
		\textcolor{blue}{Constraints} & \textcolor{blue}{[2 ; 4]} & \textcolor{blue}{10} & \textcolor{blue}{[20000 ; 30000]} & \textcolor{blue}{[8.96E+04 ; 1.01E+05]} & \textcolor{blue}{[148.51 ; 157.52]} & & &\\
		\hline
		\hline
 		\end{tabular*}
  	\label{tab_D11}
\end{table*}


\bsp	
\label{lastpage}
\end{document}